\documentclass[aip,amsmath,amssymb,reprint,onecolumn] {revtex4-1}
\usepackage[]{graphicx} 

\usepackage{dcolumn}
\usepackage{bm}

\usepackage[utf8]{inputenc}
\usepackage[T1]{fontenc}
\usepackage{mathptmx}
\usepackage{etoolbox}
\usepackage{booktabs}
\usepackage{hyperref}
\usepackage{xcolor}
\hypersetup{
  colorlinks=true, 
  linkcolor=blue,
  urlcolor=black,
  citecolor=blue,
  linkbordercolor=blue,
  urlbordercolor=cyan,
  citebordercolor=blue
} 

\makeatletter
\def\@email#1#2{%
 \endgroup
 \patchcmd{\titleblock@produce}
  {\frontmatter@RRAPformat}
  {\frontmatter@RRAPformat{\produce@RRAP{*#1\href{mailto:#2}{#2}}}\frontmatter@RRAPformat}
  {}{}
}%
\makeatother
\begin{document}

\preprint{AIP/Presumed FDF Modeling}

\title[]{Investigation of Deep Learning-Based Filtered Density Function for Large Eddy Simulation of Turbulent Scalar Mixing}
\author{Shubhangi Bansude}
\author{Reza Sheikhi}%
 \email{shubhangi.bansude@uconn.edu}
\affiliation{ 
Department of Mechanical Engineering, University of Connecticut, Storrs, CT, USA
}%
\date{\today}

\begin{abstract}

\textcolor{black}{A filtered density function (FDF) model based on deep neural network (DNN), termed DNN-FDF, is introduced for large eddy simulation (LES) of turbulent flows involving conserved scalar transport. The primary objectives of this study are to develop the DNN-FDF models and evaluate their predictive capability in accounting for various filtered moments, including that of non-linear source terms. A systematic approach is proposed to select DNN training sample size and architecture via learning curves to minimize bias and variance. Two DNN-FDF models are developed, one utilizing FDF data from Direct Numerical Simulations (DNS) of constant-density temporal mixing layer, and the other from zero-dimensional pairwise mixing stirred reactor simulations. The latter is particularly intended for cases where generating DNS data is computationally infeasible. DNN-FDF models are applied for LES of a variable-density temporal mixing layer. The accuracy and consistency of both DNN-FDF models are established by comparing their predicted filtered scalar moments with those of conventional LES, where moment transport equations are directly solved. The DNN-FDF models are shown to outperform a widely used presumed-FDF model, especially for multi-modal FDFs and higher variance values. Results are further assessed against DNS and the transported FDF method. The latter couples LES with Monte Carlo for mixture fraction FDF computation. Most importantly, the study shows that DNN-FDF models can accurately filter highly non-linear functions within variable-density flows, highlighting their potential for turbulent reacting flow simulations. Overall, the DNN-FDF approach is shown to offer an accurate yet computationally economical approach for describing turbulent scalar transport.}

\end{abstract}
\maketitle

\section{Introduction} \label{sec: introduction}
The simulation of turbulent flows remains an open challenge in spite of being the focus of intensive research for several decades. The difficulty arises due to the requirement of high spatial and temporal resolutions to compute multi-scale flow structures. The model-free approach of direct numerical simulation (DNS) entails resolving all spatiotemporal scales in the flow field. Despite the high accuracy of DNS, the exorbitant computational cost inhibits its application to flows of practical interest. Consequently, the predictive simulation methods are required to reduce the resolution requirements, while ensuring sufficient precision by accurate accounting of the unresolved features. The large eddy simulation (LES) approach facilitates the use of coarser computation grids by filtering the governing equations. In LES, only large scales of the flow field are resolved, while motions at the small scales, generally referred to as subgrid scales (SGS), are modeled. In the presence of chemical reactions, the non-linearity of chemical source terms creates an additional modeling requirement. An effective closure strategy for this purpose is to directly solve the transport equations for joint scalar filtered density function (FDF)~\cite{Givi06}. The transported FDF approach offers an exact way to formulate the filtered reaction source terms in a closed form and is significantly less intensive computationally than DNS. However, the high dimensionality of the joint FDF and the stochastic nature of the FDF approach still posit substantial computational costs for practical applications~\cite{Pope1981, Pope00, Pope85}. On the other hand, the low-fidelity moments methods are most commonly used in engineering applications due  to their relatively lower computational overhead~\cite{Girimaji1991}. One class of such methods, particularly for LES of non-premixed combustion, is based on a conserved scalar (scalar independent of chemistry). These approaches assume that the thermochemical state (species mass fractions, density, and temperature) of a system depends on mixing, hence it is a function of the extent of mixing of fuel and oxidizer. The mixing rate of fuel and oxidizer is often quantified by a conserved scalar, mixture fraction ($\phi$). The most basic form of the conserved scalar approach assumes an infinitely fast or equilibrium chemistry, making the thermochemical state a function of mixture fraction alone. Following that, the filtered value for the dependent scalar can be estimated as a weighted integral with regard to the FDF of the SGS mixture fraction. The most rigorous approach to account for the FDF of mixture fraction is by solving its transport equation. Alternatively, for ease of use and simplicity, the presumed FDF approach is introduced in which the FDF is represented by 
presumed shape functions~\cite{MdG93}. This approach despite its simplicity is still important for practical applications. An example is the non-premixed flames with infinitely fast chemistry. Although these flames are simpler than those requiring finite-rate chemistry, they still represent a large number of cases of theoretical and practical importance~\cite{jimenez1997priori}. In some approaches a conserved scalar is also carried besides accounting for finite-rate chemistry; examples include conditional moment closure (CMC)~\cite{KLIMENKO1999595}, steady and unsteady laminar flamelet models~\cite{peters1984laminar}, flamelet generated manifolds (FGM) model~\cite{oijen2000modelling}, and the closely related flamelet/progress variable (FPV) strategy~\cite{pierce2004progress}. Depending on the formulation, these methods require a presumed FDF of mixture fraction, along with that of other characteristic scalars, such as progress variables and scalar dissipation rate. 

Because the mixture fraction FDF plays a central role in many modeling strategies, its investigation is critically important to improve the accuracy of these models. Numerous established presumed FDF models are considered in the literature, including Gaussian, clipped Gaussian, Dirac $\delta$, tophat, $\beta$ function, and others~\cite{Wang2014486, Floyd2009, Cook1994}. The most prevalent among these is the $\beta$ function\cite{Cook1994} FDF. The $\beta$-FDF parameterizes the shape of FDF by the first two statistical moments and is flexible enough to approximate the behavior of mixture fraction ranging from the distribution for unmixed reactants (single or double-delta function) to that for well-mixed reactants (Gaussian)~\cite{Cook1994, Girimaji1991}. Interestingly, there is no theoretical basis to support this model~\cite{Wall2000, Wang2021} besides its direct extension of the presumed $\beta$ probability density function (PDF) of mixture fraction in Reynolds-averaged Navier-Stokes (RANS) simulations\cite{Girimaji1991}. Since filtering in LES and Reynolds-averaging in RANS lead to mathematically similar governing equations, it is a common practice to employ RANS closure strategies directly in LES. However, it is important to emphasize that the PDF in RANS essentially describes a different flow physics than the FDF in LES. The PDF in RANS describes the probability density at a fixed point sampled over infinite realizations, or in the case of statistically stationary flow, over an infinite amount of time\cite{Floyd2009}. On the other hand, the FDF in LES describes the probability of states within the filtered region for one realization at a single time\cite{Floyd2009}. Some prior studies have found that the $\beta$ function is an adequate approximation of the FDF of mixture fraction, especially with increased accuracy at the high Reynolds numbers~\cite{Cook1994, Wall2000, Yun2005, Floyd2009}. However, other computational and experimental investigations have shown that the actual FDFs can be significantly different than the $\beta$ function~\cite{jimenez1997priori, Tong1402171, Wang2021}. The experimental observations by Tong~\emph{et al.} showed the existence of bi-modal FDFs with maxima away from the bounds, which can not be represented by the $\beta$ function~\cite{Tong1402171}. Furthermore, the investigation by Floyd~\emph{et al.}~\cite{Floyd2009} and Wang~\emph{et al.}~\cite{Wang2021} have shown that the $\beta$ function is inadequate for representing multi-modal and narrow FDFs which can otherwise be captured more accurately with the transported FDF methods. 

In recent years, deep learning has been extensively employed to develop data-driven models in scientific computing. Deep learning is a specialized branch of machine learning and utilizes the deep neural network (DNN), \emph{i.e.}, artificial neural network (ANN) with more than two hidden layers. The neural network layers are embedded with non-linear activation functions that allow DNNs to approximate highly complex functions and operators. The present study aims to develop and investigate data-driven models for the FDF of the mixture fraction based on deep learning. In turbulent flows, DNNs have been applied to a multitude of modeling problems. For example, SGS stress modeling~\cite{Wang5054835}, approximation and integration of combustion chemistry~\cite{Chatzopoulos2013, bansude4142013data, Owoyele2021, Nguyen2021, Franke2017}, to construct closure models based on experimental data~\cite{Ranade2019279}, modeling of the conditional scalar dissipation rate in spray flame LES~\cite{Yao2020}. In the context of presumed FDF modeling, in a recent study, Frahan~\emph{et al.} constructed models based on three machine learning techniques (traditional methods, deep learning, and generative learning) to predict the joint FDF of mixture fraction and progress variable~\cite{T.HenrydeFrahan2019a}. All three models demonstrated improved accuracy compared to the $\beta$-FDF, the deep learning one being the fastest. Another investigation by Gitushi~\emph{et al.}~\cite{Gitushi2022}, proposed a hybrid PDF-like framework where a deep operator network (DeepONet) is employed to construct pointwise joint PDF of principal components~\cite{Gitushi2022}. The work by Yao~\emph{et al.}~\cite{Yao2020} presented the application of DNN to model FDF of mixture fraction in carrier-phase LES of turbulent spray combustion~\cite{Yao2020}. 

\textcolor{black}{The present study aims to further leverage DNN for modeling of presumed FDF of mixture fraction. Specifically, we develop and evaluate the DNN-based FDF model, termed DNN-FDF, in the context of turbulent flows. These models are then applied to the LES of a variable-density, three-dimensional (3-D), temporal mixing layer characterized by conserved scalar mixing. Our primary objectives encompass four key aspects. First, we propose a systematic method based on learning curves to select the training sample size and network architecture, which has not been adequately addressed in previous studies. Second, we explore the development of DNN-FDF models with two distinct training data sources: (1) a model trained on FDF data derived from Direct Numerical Simulations (DNS) of a temporally evolving constant-density mixing layer, and (2) a model trained on FDFs synthesized from simulations within a zero-dimensional (0-D) pairwise mixing stirred reactor (PMSR). Exploring PMSR as an alternative way of training data generation is significant, given the prohibitive computational costs associated with DNS in practical applications. Third, we conduct extensive investigations into the consistency and convergence of DNN-FDF models when applied to LES of turbulent mixing layer by rigorous comparative analysis against DNS, transported FDF, and the widely-used presumed $\beta$-FDF approach. Finally, we demonstrate the potential of DNN-FDF models for predicting turbulent reactive flows by addressing essential factors such as their adaptability to variable-density mixing layers, capability to filter non-linear functions, and their performance using different filter and grid sizes.}

The paper is organized as follows: Section~\ref{sec: formulation} outlines the formulation of DNS and LES-FDF simulations. 
Section~\ref{sec: DNN_models} provides an overview of DNN models for FDF prediction including training data generation, network architecture selection, and model validations. The subsequent section~\ref{sec: results_mixing_layer} discusses the applications of DNN-FDF models to temporally evolving 3-D mixing layers. Finally, Section~\ref{sec: conclusions} provides the concluding remarks along with an overview of the capabilities of the proposed methodology.
\section{Formulation} \label{sec: formulation}
In variable density turbulent flows with passive scalar transport, the primary dependent variables are density ($\rho$), velocity vector ($u_{i}$) in $x_{i}$ direction for $i=1,2,3$, pressure ($p$), and the mixture fraction ($\phi$). The transport equations that govern these variables include the continuity, conservation of momentum, and passive scalar transport equations
\begin{equation} 
\begin{split}
& \frac{\partial \rho}{\partial t}+\frac{\partial \rho u_{i}}{\partial x_{i}} = 0, \\
& \frac{\partial \rho u_{j}}{\partial t}+\frac{\partial \rho u_{i} u_{j}}{\partial x_{i}} =-\frac{\partial p}{\partial x_{j}}+\frac{\partial \tau_{i j}}{\partial x_{i}}, \\
& \frac{\partial \rho \phi}{\partial t}+\frac{\partial \rho u_{i} \phi}{\partial x_{i}} =-\frac{\partial J_{i}^{}}{\partial x_{i}}
\label{Governing_equation}
\end{split}
\end{equation}
along with the ideal gas equation of state $p =\rho RT$. In these equations, $t$ represents time, $\tau_{i j}$ denotes a viscous stress tensor, $J_{i}$ denotes the scalar flux, $T$ is the temperature and $R$ is the mixture gas constant. For a Newtonian fluid with Fick's law of diffusion, the $\tau_{i j}$ and $J_{i}$ 
are represented by Eq.(\ref{tau_ij_flux})
\begin{equation} 
\begin{split}
& \tau_{i j}=\mu\left(\frac{\partial u_{i}}{\partial x_{j}}+\frac{\partial u_{j}}{\partial x_{i}}-\frac{2}{3} \frac{\partial u_{k}}{\partial x_{k}} \delta_{i j}\right), \\
& J_{i}=-\gamma \frac{\partial \phi}{\partial x_{i}}
\label{tau_ij_flux}
\end{split}
\end{equation}
where $\mu$ is the dynamic viscosity and $\gamma=\rho \Gamma$ denotes the the mass molecular diffusivity coefficient. Both $\mu$ and $\gamma$ are assumed constant and the Lewis number is assumed to be unity~\cite{Jaberi1999, Sheikhi2007}.

Large eddy simulation involves the spatial filtering of Eq.~(\ref{Governing_equation}) with operation mathematically described as 
\begin{equation} 
\langle f(\mathbf{x}, t)\rangle_{\ell}=\int_{-\infty}^{+\infty} f\left(\mathbf{x}^{\prime}, t\right) G\left(\mathbf{x}^{\prime}, \mathbf{x}\right) d \mathbf{x}^{\prime}
\end{equation}
where $G\left(\mathbf{x}^{\prime}, \mathbf{x}\right)$ denotes a filter function, and $\langle f(\mathbf{x}, t)\rangle_{\ell}$ is the filtered value of the transport variable $f(\mathbf{x}, t)$. In a variable density flows, it is convenient to use the Favre-filtered quantity $\langle f(\mathbf{x}, t)\rangle_L=\langle\rho f\rangle_{\ell} /\langle\rho\rangle_{\ell}$. We consider a filter function with characteristic width $\Delta_f$ that is spatially and temporally invariant and localized, thus $G\left(\mathbf{x}^{\prime}, \mathbf{x}\right) \equiv G\left(\mathbf{x}^{\prime}-\mathbf{x}\right)$ with the properties $G(\mathbf{x}) \geq 0$, and $\int_{-\infty}^{\infty} G(\mathbf{x}) \mathrm{d} \mathbf{x}=1$~\cite{Jaberi1999, Sheikhi2007}. The application of the filtering operation to the transport Eq.~(\ref{Governing_equation}) yields
\begin{equation} 
\begin{split}
& \frac{\partial\langle\rho\rangle_{\ell}}{\partial t}+\frac{\partial\langle\rho\rangle_{\ell}\left\langle u_{i}\right\rangle_{L}}{\partial x_{i}}=0, \\
&\frac{\partial\langle\rho\rangle_{\ell}\left\langle u_{j}\right\rangle_{L}}{\partial t}+\frac{\partial\langle\rho\rangle_{\ell}\left\langle u_{i}\right\rangle_{L}\left\langle u_{j}\right\rangle_{L}}{\partial x_{i}}=-\frac{\partial\langle p\rangle_{\ell}}{\partial x_{j}}+\frac{\partial\left\langle\tau_{i j}\right\rangle_{\ell}}{\partial x_{i}}-\frac{\partial T_{i j}}{\partial x_{i}}, \\
&\frac{\partial\langle\rho\rangle_{\ell}\left\langle\phi\right\rangle_{L}}{\partial t}+\frac{\partial\langle\rho\rangle_{\ell}\left\langle u_{i}\right\rangle_{L}\left\langle\phi\right\rangle_{L}}{\partial x_{i}}=-\frac{\partial\left\langle J_{i}^{\alpha}\right\rangle_{\ell}}{\partial x_{i}}-\frac{\partial M_{i}^{\alpha}}{\partial x_{i}}
\label{LES_filterd_equations}
\end{split}
\end{equation}
where $T_{i j}=\langle\rho\rangle_{\ell}\left(\left\langle u_{i} u_{j}\right\rangle_{L}-\left\langle u_{i}\right\rangle_{L}\left\langle u_{j}\right\rangle_{L}\right)$ and $M_{i}^{\alpha}=\langle\rho\rangle_{\ell}\left(\left\langle u_{i} \phi\right\rangle_{L}-\left\langle u_{i}\right\rangle_{L}\left\langle\phi\right\rangle_{L}\right)$ denote the SGS stress and the SGS scalar flux, respectively. We adopt the Smagorinsky closure model for $T_{i j}$ and $M_{i}^{\alpha}$, rendering the filtered scalar transport equation as
\begin{equation}
\frac{\partial\left(\langle\rho\rangle_{\ell}\left\langle\phi\right\rangle_{L}\right)}{\partial t}+\frac{\partial\left(\langle\rho\rangle_{\ell}\left\langle u_{i}\right\rangle_{L}\left\langle\phi\right\rangle_{L}\right)}{\partial x_{i}}=\frac{\partial}{\partial x_{i}}\left[\left(\gamma+\gamma_t\right) \frac{\partial\left\langle\phi\right\rangle_{L}}{\partial x_{i}}\right]
\label{filterd_mean_equations}
\end{equation}
where the SGS diffusivity coefficient $\gamma_{t}=\langle\rho\rangle_{\ell} \Gamma_{t}$ in which $\Gamma_{t}=\nu_{t} / S c_{t}$; $\nu_t$ is the SGS viscosity and $S c_{t}$ is the SGS Schmidt number which is assumed to be constant and equal to the SGS Prandtl number~\cite{Colucci1998, Jaberi1999, Sheikhi2007}. 

The complete information about the statistical variation of $\phi$ within the SGS is contained in the scalar FDF, denoted by $F_L(\psi;\mathbf{x},t)$, where $\psi$ is the sample space variable corresponding to $\phi$. Therefore, with the knowledge of the FDF, the filtered form of any function $Q(\phi)$ of the scalar, can be obtained as~\cite{Colucci1998, Jaberi1999}
\begin{equation}
\langle \rho \rangle_{\ell} \langle Q\rangle_{L} =\int_{0}^{1} Q(\psi) F_L(\psi;\mathbf{x},t) d \psi
\label{TPDF_favre_filtering}
\end{equation}
Two methods are typically used to determine the FDF~\cite{DSMG07, Pitsch06, Haworth10}: (i) the presumed FDF, and (ii) the transported FDF methods. \textcolor{black}{In the presumed FDF method, the FDF of the SGS variables is specified {\it a priori}~\cite{MdG93, Cook1994, Wall2000}. The parameters of such FDF are determined using the moments calculated from their respective transport equations. The most widely used approach of the $\beta$-FDF, denoted by $P\left(\psi; \left\langle\phi\right\rangle_{L}, \sigma_\phi \right)$, parameterizes the presumed FDF using the first and second scalar moments of $\phi$: $\left\langle\phi\right\rangle_{L}$ and $\sigma_\phi=\left\langle \phi \phi\right\rangle_{L}-\left\langle \phi \right\rangle_{L}\left\langle \phi \right\rangle_{L}$, given as 
\begin{equation} 
\begin{split}
\label{beta_function} & P\left(\psi; \left\langle\phi\right\rangle_{L}, \sigma_\phi \right)= \frac{\psi^{a-1}(1-\psi)^{b-1} \Gamma(a+b)}{\Gamma(a) \Gamma(b)}  \\ &  a =\langle \phi \rangle_L\left(\frac{\langle \phi \rangle_L(1-\langle \phi \rangle_L)}{\sigma_\phi}-1\right), \quad b=\left (\frac{a}{\langle \phi \rangle_L} -a \right)
\end{split}
\end{equation}  
where $\Gamma$ represents the standard gamma function. The DNN-FDF models developed in the present study also utilize mean and variance as inputs, providing a presumed FDF model based on DNN. Once the FDF is obtained using the $\beta$-FDF or DNN-FDF, the resulting scalar moments are obtained using Eq.~(\ref{TPDF_favre_filtering}) with $F_L(\psi;\mathbf{x},t) \equiv \langle \rho \rangle_{\ell} P\left(\psi; \left\langle\phi\right\rangle_{L}, \sigma_\phi \right)$.}

\textcolor{black}{It is important to note that the choice of employing mean and variance as inputs in DNN-FDF is guided by two fundamental reasons. Firstly, this approach is firmly rooted in the widely accepted practice of constructing presumed FDF at least based on their first and second moments, which has been demonstrated to provide reasonable results in turbulent combustion simulations~\cite{Pitsch06, Girimaji1991, Chen19}. Besides, comparing DNN-FDF models with the $\beta$-FDF, which also relies on the first two moments, ensures a consistent and appropriate benchmark for evaluations of DNN-FDF. Secondly, employing only the mean and variance as inputs allows effective capturing of the essential statistical properties of the FDF while maintaining the computational efficiency. Including higher-order moments (or any other parameters that are not available from the resolved fields) as inputs would introduce additional transport equations, exacerbating the closure problem in turbulent combustion models and increasing computational costs. Besides, accurate representation of higher-order moments becomes increasingly sensitive to numerical errors and uncertainties in model assumptions, posing additional challenges in their implementation.}

In the transported FDF approach~\cite{Colucci1998, Jaberi1999, Sheikhi2007} the FDF is obtained from its transport equation, which is represented by a set of stochastic differential equations (SDEs)
\begin{align}
\label{particle_position_eqn} & \mathrm{d} x^{+}_i(t)=\left(\left\langle u_{i}\right\rangle_{L}+\frac{1}{\langle\rho\rangle_{\ell}} \frac{\partial\gamma+\gamma_{t}}{\partial x_{i}}\right)\mathrm{d} t+\sqrt{\frac{2\left(\gamma+\gamma_{t}\right)}{\langle\rho\rangle_{\ell}}}\,\mathrm{d} W_{i}(t) \\
\label{particle_scalar_eqn}  & \mathrm{d} \phi^{+}(t)=-\Omega_{m}\left(\phi^{+}(t)-\left\langle\phi\right\rangle_{L}\right){\mathrm{d} t}
\end{align}    
where $W_{i}$ is the Wiener process~\cite{Gardiner90} and $x^{+}_{i}(t)$ ($i=1,2,3$) and $\phi^{+}(t)$ denote the stochastic processes corresponding to the position vector and the conserved scalar variable (mixture fraction). Equation~(\ref{particle_scalar_eqn}) describes the conserved scalar variation due to turbulent mixing at the SGS which is modeled using the linear mean-square estimation (LMSE) model \cite{OBrien80, DO76}. This model includes mixing frequency ($\Omega_{m}$) within the SGS which is modeled as $\Omega_{m}=C_{\Omega}\left(\gamma+\gamma_{t}\right) /\left(\langle\rho\rangle_{\ell} \Delta^{2}\right)$, where $C_{\Omega}$ is a model constant. As detailed in Ref.~\cite{Colucci1998, Jaberi1999}, transport equations implied by the system of SDEs (Eqs.~(\ref{particle_position_eqn},\ref{particle_scalar_eqn})) can be derived for various filtered moments. The first order moment is the transport of $\left\langle\phi\right\rangle_{L}$ which is consistent with Eq.~(\ref{filterd_mean_equations}). The second order moment is that of 
the scalar SGS variance
\begin{equation}
\frac{\partial \langle\rho\rangle_{\ell}\, \sigma_\phi}{\partial t}+\frac{\partial \langle\rho\rangle_{\ell}\left\langle u_{i}\right\rangle_{L}\,\sigma_\phi}{\partial x_{i}}=\frac{\partial}{\partial x_{i}}\left[\left(\gamma+\gamma_{t}\right) \frac{\partial \sigma_\phi}{\partial x_{i}}\right]+2\left(\gamma+\gamma_{t}\right)\left[\frac{\partial\left\langle\phi\right\rangle_{L}}{\partial x_{i}} \frac{\partial\left\langle\phi\right\rangle_{L}}{\partial x_{i}}\right]-2 \Omega_{m}\langle\rho\rangle_{\ell} \sigma_\phi
\label{variance_equations}
\end{equation}
wherein the last term on the right hand side is the scalar dissipation described by the LMSE model. \textcolor{black}{This equation is adopted here consistent with the transported FDF approach~\cite{Sheikhi2007,Jaberi1999}. The purpose is to provide a comparison between the DNN-FDF and transported FDF results, as presented below. It is clear that this choice does not in any way limit the DNN-FDF results and in general, any other model can be used instead to obtain the variance.}

The numerical solution approach to obtain the transported FDF is a hybrid finite-difference/Monte Carlo procedure in which the finite-difference (FD) method is used to solve the filtered transport equations Eqs.~(\ref{LES_filterd_equations}, \ref{filterd_mean_equations}) while the system of SDEs is solved by the Lagrangian Monte Carlo (MC) method. The latter provides a representation of the FDF using an ensemble of $N_p$ MC particles carrying the information about their position in space, $x_i^{(n)}(t)$, and scalar values, $\phi^{(n)}(t)$ where $n=1,\ldots, N_p$. This information is updated via temporal integration of the SDEs. The computational domain is discretized on equally-spaced FD grid points. The statistical information from the MC solver is obtained by considering an ensemble of $N_E$ computational particles residing within an ensemble domain of characteristic width $\Delta_E$ centered around each grid point. For reliable statistics with minimal numerical error, it is desired to minimize the size of an ensemble domain and maximize the number of MC particles inside it. This causes convergence of the ensemble averaged statistics to the desired filtered quantities. Similar to previous studies~\cite{Colucci1998, Jaberi1999, Sheikhi2007}, in order to reduce the computational cost, MC particles have non-uniform weights. This procedure allows a smaller number of particles in regions where a low degree of variability is expected. It has been shown~\cite{Colucci1998, Jaberi1999} that the sum of weights within the ensemble domain is related to filtered fluid density as
\begin{equation}
\langle \rho \rangle_\ell \approx \frac{\Delta m}{V_E} \sum_{{n \in
\Delta_E} }w^{(n)}    
\label{EQ:rho_FDF}
\end{equation}
where $V_E$ is
the volume of the ensemble domain and $\Delta m$ is the particle mass with unit weight. The Favre-filtered value of any quantity $\langle Q\rangle_{L} $ is constructed at each FD grid point as
\begin{equation}
\langle Q\rangle_{L} \approx \frac{\sum_{n \in \Delta_{E}} w^{(n)} \hat{Q}\left(\phi^{(n)}\right)}{\sum_{n \in \Delta_{E}} w^{(n)}}
\label{Q_ensemble_mean}
\end{equation}
The FD solver involves a compact parameter finite-difference scheme which is a variant of the MacCormack scheme with fourth-order spatial accuracy as well as a second-order predictor-corrector sequence for time discretization. The transfer of information from the grid points to the MC particles is done using linear interpolation. The FD solver provides the variables needed for solving the SDEs such as the velocity field. The first two scalar moments can be obtained from the MC solver according to Eq.~(\ref{Q_ensemble_mean}) as well as the FD solver by solving Eqs.~(\ref{LES_filterd_equations}, \ref{variance_equations}) by FD method. As shown in previous studies~\cite{Colucci1998, Jaberi1999, Sheikhi2007}, such redundancy is quite useful for monitoring the accuracy of the results obtained from both solvers.
For more information about the transported scalar FDF methodology, we refer to previous work~\cite{Colucci1998, Jaberi1999}.

In the present study, three simulation approaches are employed for LES simulation. 
\begin{enumerate}
    \item In the first approach, which is the primary objective of this study and referred to as LES-FD, the LES transport equations Eqs.~(\ref{LES_filterd_equations}) along with those of the first two scalar moments, Eqs.~(\ref{filterd_mean_equations}, \ref{variance_equations}) are solved using the FD method. In these simulations, the DNN-FDF provides the presumed FDF. 
    \item The second approach is similar to the previous one except that the presumed FDF is described by the $\beta$ distribution. The purpose of these simulations is to evaluate the accuracy of the DNN-FDF in conjunction with the conventional presumed FDF approach~\cite{MdG93, Cook1994, Wall2000}. 
    \item To further assess the performance of the DNN-FDF approach, we conduct simulations using the transported scalar FDF methodology. These simulations, referred to as LES-MC, are based on the hybrid FD/MC procedure as described above. This approach provides a faithful representation of the evolution of the scalar FDF along with its filtered moments which is instrumental for appraising the DNN-FDF approach.
\end{enumerate}
\section{Deep Neural Network for FDF prediction } \label{sec: DNN_models}
\subsection{Training Data Generation} \label{subsec: PPDF_data}
A sample dataset of FDFs is generated using the DNS data of constant density 3-D temporal mixing layer at different times. One sample is defined as a pointwise sampling of the FDF along with a corresponding set of moments, which constitute the filtered Favre mean and the SGS Favre variance of the mixture fraction. The sample moments are generated using a discrete box filter described by
\begin{equation}
\langle{\varphi}(x, y, z)\rangle=\frac{1}{N_f^3} \sum_{i=-N_f / 2}^{N_f / 2}\ \sum_{j=-N_f / 2}^{N_f / 2}\ \sum_{k=-N_f / 2}^{N_f / 2} \varphi(x+i \Delta, y+j \Delta, z+k \Delta)
\label{DNS_filtering_eqn}
\end{equation}
where $\varphi$ represents any variable to be filtered and $\langle{\varphi}(x, y, z)\rangle$ denotes the filtered value of $\varphi$; $N_f=12$ is the number of points used for filtering in each direction and $\Delta=\Delta x=\Delta y=\Delta z$ is DNS spatial step size--- the box thus has a size of $6 \Delta$ in all coordinate directions representing a filter size of ${\Delta_f}=12 \Delta$. This filter size is chosen consistent with previous studies~\cite{Sheikhi2003, Sheikhi2007} and ensures adequate sampling to construct the FDF along with its first two moments. Alongside the filtered moments, the FDF of $\phi$, $F_L(\psi; \mathbf{x}, t)$, is directly constructed from the DNS data by binning $\phi$ at $N_f^3$ grid points within the filter domain, as defined above, into equally spaced bins from $\phi=0$ to $\phi=1$. In this study, $32$ such bins are chosen to construct the FDFs; any number of bins resulting in a well-defined distribution may however be specified.

\begin{figure}[htbp]
\centerline{\includegraphics[width = \columnwidth]{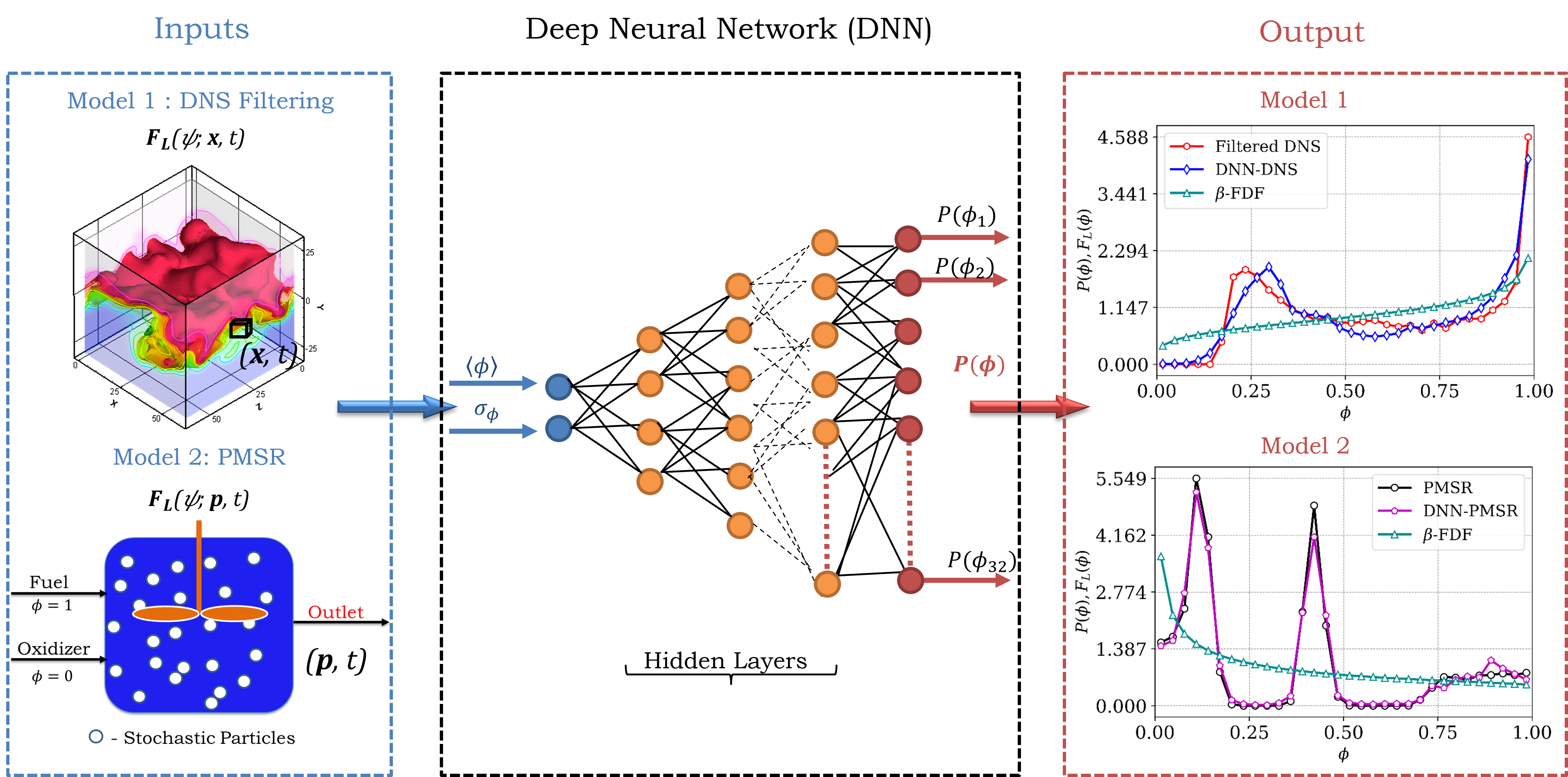}}
\caption{Graphical representation of the process of generating training data, designing a network architecture, and training for a Deep Neural Network (DNN) to predict the FDF.}
\label{PPDF_workflow}
\end{figure}

In addition to DNS data, in this study we introduce generating training data using a zero-dimensional PMSR. As a part of this study, we evaluate this approach which may serve as an alternative means of training the DNN for cases wherein DNS data is not available. This approach is similar to utilizing the DNS data, except that the spatial averaging in Eq.~(\ref{DNS_filtering_eqn}) is replaced by ensemble averaging over an ensemble of $N$ notional particles within the PMSR,
\begin{equation}
\langle{\varphi}\rangle=\frac{1}{N} \sum_{n=1}^{N} \varphi^{(n)}
\label{PMSR_filtering_eqn}
\end{equation}
where $\varphi^{(n)}$ denotes the values of $\varphi$ carried by particle $n$ and $\langle{\varphi}\rangle$ denoted its ensemble-averaged value. The merit of this approach is due to the capacity of the PMSR to provide a plausible representation of the events occurring at a single computational cell in actual turbulent (reactive) flow simulations, as pointed out in several studies ~\cite{yang1998investigation, singer2006, Ranade2019279, hadi2016study, Bansude2023}. In PMSR, the mixing process is modeled as a combination of macro-mixing and micro-mixing. Macro-mixing, with its residence timescale ($\tau_r$) and pairing timescale ($\tau_p$), refers to large-scale mixing events caused by fluid particle movement. Micro-mixing, on the other hand, is molecular-scale mixing, characterized by a mixing timescale ($\tau_m$). PMSR exhibits ideal macro-mixing but imperfect micro-mixing. The reactor at any given time step $t$, is composed of an even number $N$ of particle, initially arranged in pairs $(p,q)$ such that the particles $(1, 2), (3, 4), ...,(N-1, N)$ are partners. At each discrete time step $dt$, three events occur inflow, outflow, and pairing, which change the composition of $n^{th}$ particle, ${\phi}^{(n)}(t)$. The inflow and outflow events involve randomly selecting $\frac{dt}{\tau_r} \frac{N}{2}$ pairs and replacing their $\phi$ with the inlet stream. The inlet streams consist of fuel (represented by $\phi=1$) and oxidizer (represented by $\phi=0$) streams. The pairing event involves randomly selecting $\frac{dt}{\tau_p} \frac{N}{2}$ particle pairs, different from the inflow particles, and shuffling them to alter their compositions. Between these discrete times, particle pairs $(p,q)$ evolve through mixing as
\begin{equation}
\begin{split}
\frac{d{\phi}^{(p)}}{dt} = \frac{({\phi}^{(p)}-{\phi}^{(q)})}{\tau_m}\\
\frac{d{\phi}^{(q)}}{dt} = \frac{({\phi}^{(q)}-{\phi}^{(p)})}{\tau_m}
\label{pmsr_mixing_step}
\end{split}
\end{equation}
The representative training dataset of the mixing layer is generated by sampling FDFs from multiple PMSR simulations with varying residence, paring, and mixing timescales. Each simulation carries $10^4$ particles. The generation of FDF, $F_L(\psi; p, t)$, follows the same procedure as sampling from the DNS data, as explained above, except that the sample points are comprised of instantaneous data from $N$ particles at each timestep as depicted in Fig.~(\ref{PPDF_workflow}).

\begin{figure}[htbp]
\centerline{\includegraphics[width = \columnwidth]{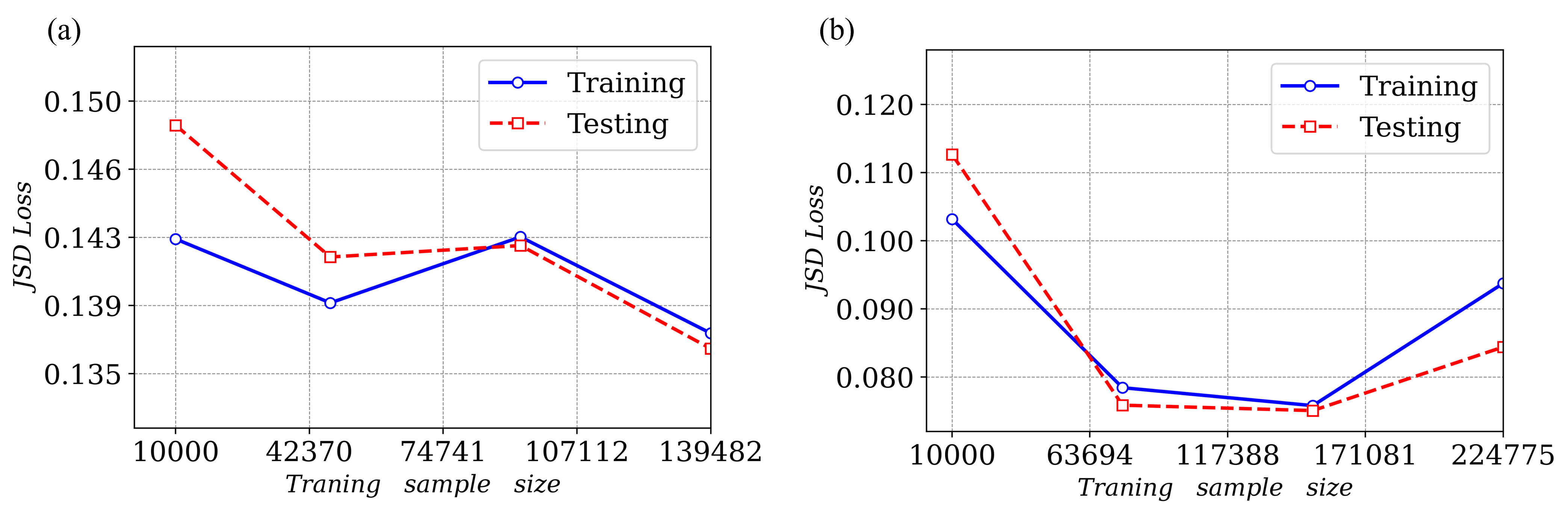}}
\caption{Learning curves depicting mean training and testing error obtained from converged models trained on progressively larger sample sizes: (a) DNN-DNS model, and (b) DNN-PMSR model.}
\label{Learning_curves}
\end{figure}
\begin{figure}[htbp]
\centerline{\includegraphics[width = \columnwidth]{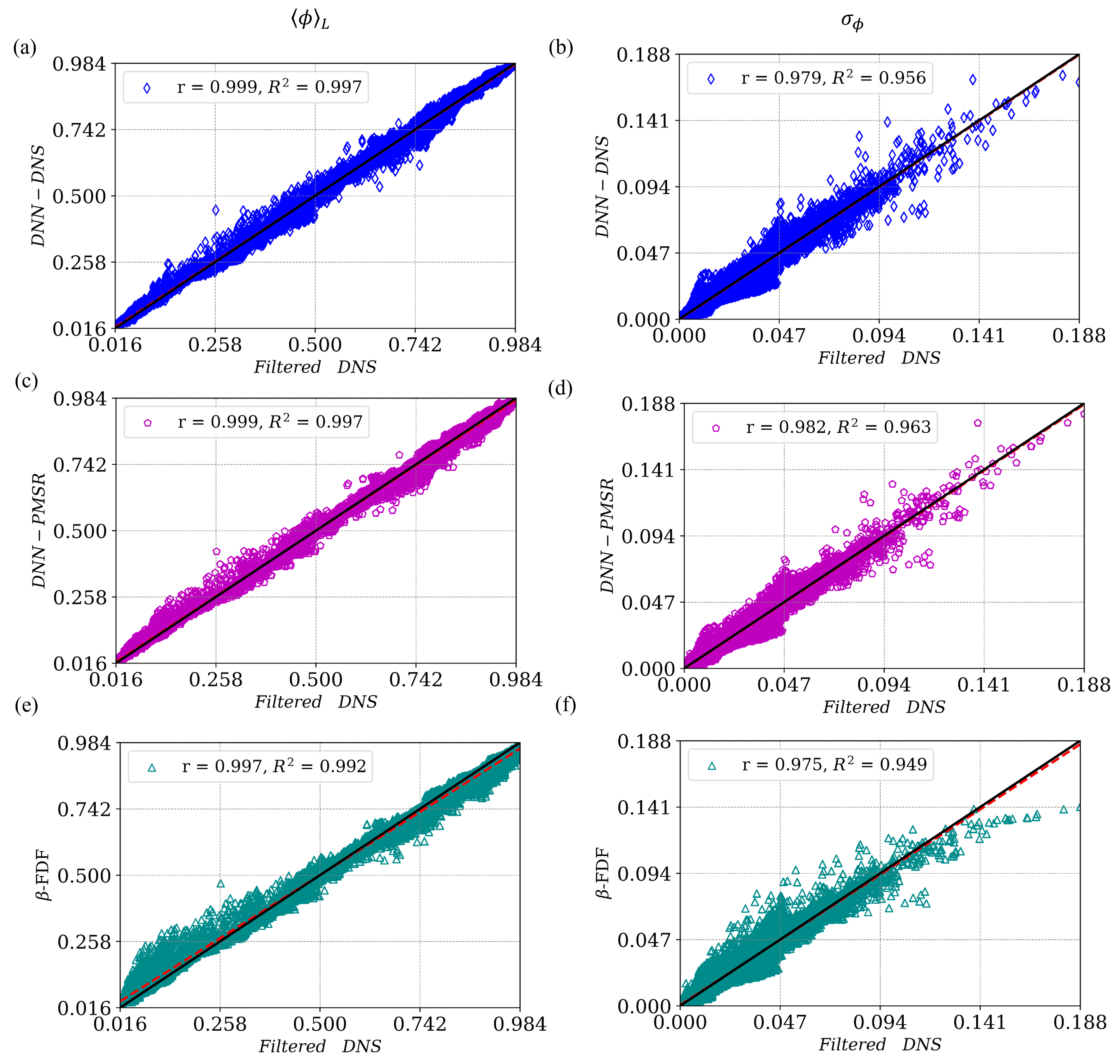}}
\caption{\textcolor{black}{Scatter plots of scalar statistics ${\langle\phi\rangle_L}$ (column 1) and ${\sigma_{\phi}}$ (column 2) for a validation set generated from constant density mixing layer DNS. Comparison of filtered DNS with (a, b) DNN-DNS model, (c, d) DNN-PMSR model (e, f) $\beta$-FDF model. The solid and dashed lines denote the linear regression and $45^\circ$ lines, respectively. $r$ denotes the correlation coefficient, $R^2$ denotes the coefficient of determination.}}
\label{DNN_validation_Moments_prediction}
\end{figure}

\begin{figure}[htbp]
\centerline{\includegraphics[width = 0.5\columnwidth]{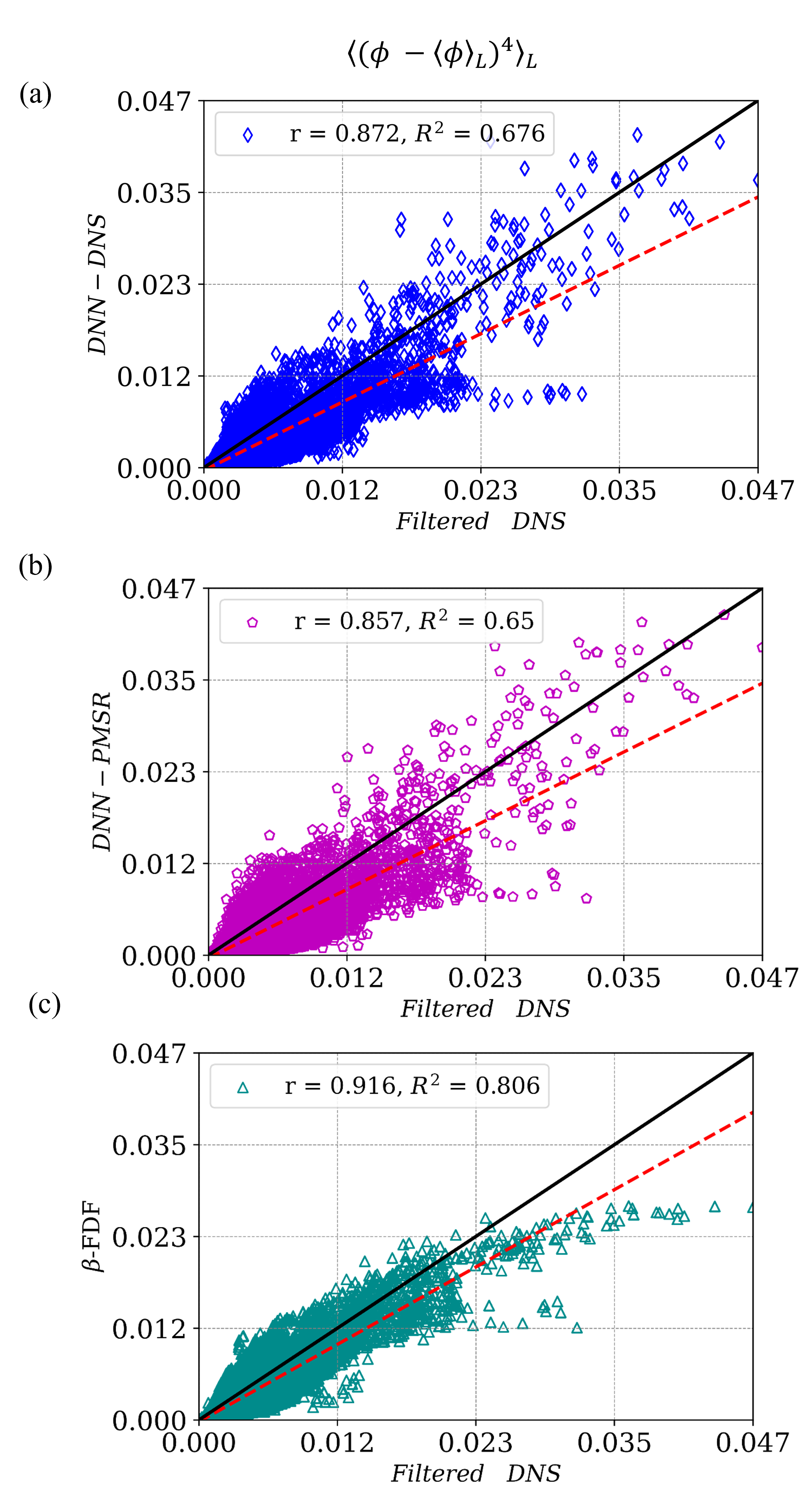}}
\caption{\textcolor{black}{Scatter plots of the fourth central scalar moment, $\left\langle \left(\phi-\left\langle \phi \right\rangle_{L} \right)^{4}\right\rangle_{L}$, for a validation set generated from constant density mixing layer DNS. Comparison of filtered DNS with (a) DNN-DNS model, (b) DNN-PMSR model (c) $\beta$-FDF model. The solid and dashed lines denote the linear regression and $45^\circ$ lines, respectively. $r$ denotes the correlation coefficient, $R^2$ denotes the coefficient of determination.}}
\label{DNN_validation_fourth_Moments_prediction}
\end{figure}

\begin{figure}[htbp] 
\centerline{\includegraphics[width = \columnwidth]{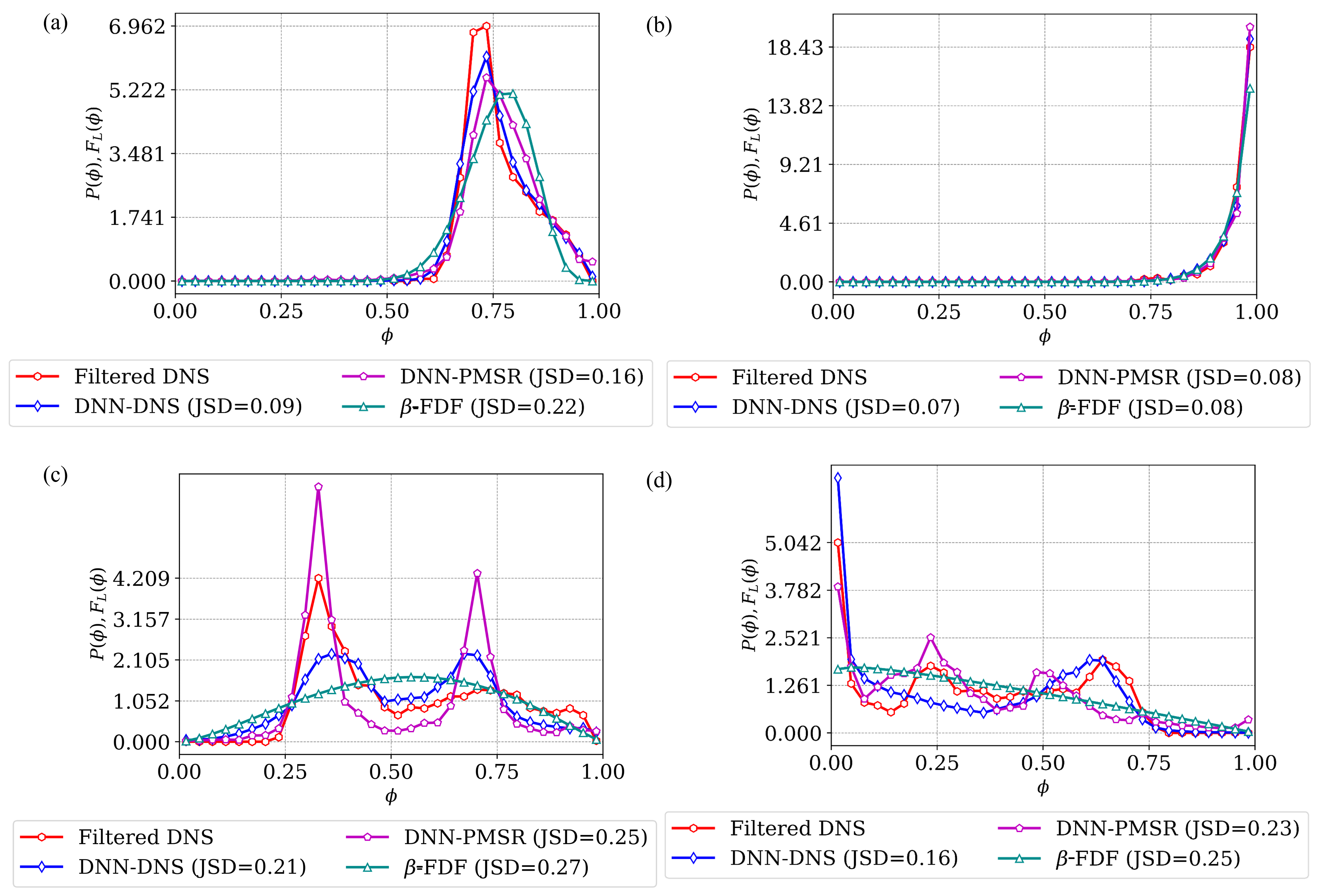}}
\caption{(a-d) Sample FDFs obtained for randomly selected ($\langle \phi \rangle_L$, $\sigma_{\phi}$).  The legends indicate the degree of similarity between the FDFs predicted by the model and the corresponding filtered FDFs obtained from DNS, as measured by the Jensen-Shannon Divergence (JSD).}
\label{DNN_validation_Sample_FDF}
\end{figure}
\subsection{DNN Architecture Selection and Training} \label{subsec:PPDF_training}
The problem of modeling the FDF can be interpreted as a multi-output classification problem, for which we adopted a fully connected multi-layer perceptron (MLP) network. The DNN model takes the first two statistical moments of $\phi$ as inputs and predicts the presumed FDF, $P\left(\psi; \left\langle\phi\right\rangle_{L}, \sigma_\phi \right)$, as the output. The neural network architecture is designed to resemble a decoder network, similar to those used in encoder-decoder models in image processing which transform a compressed low-dimensional representation into a high-dimensional image representation~\cite{Goodfellow2018}. The neural network consists of $8$ hidden layers with a total of $13252$ learnable parameters. Deeper networks with fewer neurons are selected over broader networks since the increasing number of neurons increases the complexity and evaluation time proportionately. For instance, the DNN architecture employed in the work of Frahan~\emph{et al.}~\cite{T.HenrydeFrahan2019a} has two hidden layers with $256$ and $512$ neurons, respectively, and resulted in  1.1 million learnable parameters. Each hidden layer in the present DNN comprises of between $2$ to $64$ fully connected neurons followed by a leaky rectified linear unit (ReLU) activation function and a batch normalization layer, respectively. The ReLU activation function is chosen to mitigate potential issues with vanishing gradients of loss function caused by certain activation functions (\emph{e.g.}, tanh and sigmoid functions) making the network hard to train. Batch normalization of intermediate hidden layer distributions allows for smoother gradients for faster training and more accurate generalization. Lastly, to predict the FDF, we applied a softmax activation function to the outermost layer 
\begin{equation}
\boldsymbol{y}=S(\boldsymbol{x})=\frac{\exp (\boldsymbol{x})}{\sum_{i=1}^n \exp \left(\boldsymbol{x}_i\right)}
\label{softmax}
\end{equation}
where $\boldsymbol{x}$ and  $\boldsymbol{y}$ denote the layer input and output vectors of size $n$, respectively. The softmax function ensures that the predicted output has the required properties~, \emph{i.e.}, $\sum_{i=1}^n \boldsymbol{y}_i=1$ and $\boldsymbol{y}_i\in[0,1]\, \forall i=1, \ldots, n$ such that the output multiplied by the number of bins provides the actual probability densities. Finally, the binary cross entropy (BCE) is selected to measure the loss between the target ($\boldsymbol{y_t}$) and the predicted output ($\boldsymbol{y}$) 
\begin{equation}
l(\boldsymbol{y}, \boldsymbol{y_t})=\frac{1}{n} \sum_{i=1}^n\left[\boldsymbol{y}_{\boldsymbol{t}_i} \log \left(\boldsymbol{y}_i\right)+\left(1-\boldsymbol{y}_{\boldsymbol{t}_i}\right) \log \left(1-\boldsymbol{y}_i\right)\right]
\label{BCE_loaa}
\end{equation}
The training process involves $2000$ epochs until the convergence of the training loss. During each epoch, a forward pass is conducted to calculate the loss of the entire training dataset, followed by backpropagation to calculate the derivatives of the loss function and update the network parameters. The gradient descent algorithm utilized is the Adam optimizer with a learning rate ranging from $10^{-3}$ to $10^{-4}$. Two models are trained, one using data generated from DNS (DNN-DNS) and the other using data generated from PMSR (DNN-PMSR). \textcolor{black}{In both models, the respective training data is subject to random shuffling and partitioned into batches of $64$ samples each, with an allocation of $40\%$ of the entire dataset for testing purposes. Furthermore, the testing dataset associated with the DNN-DNS model, which comprises $10^5$ samples, is also employed as a validation dataset for further elaborate cross-validation of both models, as described in Section~\ref{subsec: DNN validation}.} The training process of each DNN network is illustrated in Fig.(\ref{PPDF_workflow}). It should be noted that the mathematical foundations and internal mechanisms of DNN components and training algorithms are not the focus of this study and readers are referred to Ref.~\cite{NEURIPS2019_9015} for more details. 

One of the challenges in building machine learning models is determining the optimal number of training samples that minimize the bias and variance in predictions. Bias, characterized by high values of training and testing errors, can be reduced by increasing the complexity of the model. Conversely, variance (also referred to as the generalization gap) is defined as the discrepancy between training and testing errors and can be reduced by incorporating more training examples. To determine the optimal number of training samples, we use the learning curve method~\cite{Goodfellow2018}. The learning curve indicates the relationship between the training and testing errors as the number of training samples ($N_{s}$) increases. For each model, the training data is increased from $10^4$ to the total available training sample size, and converged models are obtained for each $N_s$. The converged models are then evaluated on the corresponding training and the previously reserved common validation dataset. The errors are quantified using the mean Jensen-Shannon divergence (JSD) with lower JSD values indicating greater similarity between the target and predicted FDF. The JSD between two probability vectors $p$ and $q$ is mathematically defined as:
\begin{equation}
\begin{split}   
&JSD(p,q)= \sqrt{\frac{D(p \| m)+D(q \| m)}{2}} \\
&D(p \| m)= \begin{cases}p \log (p / m)-p+m & p>0, m>0 \\ m & p=0, m \geq 0 \\ \infty & \text { otherwise }\end{cases}
\end{split}
\label{JSD}
\end{equation}
where $m$ is the pointwise mean of $p$ and $q$, and $D$ is the Kullback-Leibler divergence~\cite{2020SciPy-NMeth}. By analyzing the learning curve for each model and selecting the optimal number of training samples, we can minimize bias and variance and improve the performance of the DNN-FDF model. As shown in Fig.~(\ref{Learning_curves}), we observe that for models DNN-DNS and DNN-PMSR, the mean JSD of the training data is lower than the testing data when $N_{s}=10^4$, indicating higher variance in the predictions. Additionally, the JSD loss values for both the training and testing datasets are higher than the ideal model loss of JSD=$0$, which suggests a higher bias. This indicates that the training dataset with $N_{s}=10^4$ is not representative enough to accurately learn the problem compared to the testing dataset used for evaluation. Similar trends are observed for the loss curves with respect to the epochs for the complete training process, which is not shown here for brevity. 

For the DNN-FDF model, further increasing the training sample size results in reduced mean training and testing errors and also, a smaller difference between them. The optimal model is identified by a training and testing loss that decreases to the point of stability with a minimal generalization gap. We achieve this for models with $N_{s} > 9\times 10^4$. Thus, we select a model with $N_{s}=1.4\times10^5$ as the suitable model for further FDF testing, as it has the lowest bias and variance. On the other hand, the learning curve of the DNN-PMSR model exhibits a slightly different trend, with training and testing errors decreasing as $N_s$ increases while the generalization gap remains low throughout. This behavior may be attributed to the PMSR dataset containing a larger set of statistically similar FDFs compared to DNS data such that it balances the testing and training datasets well even with smaller sample sizes. The model achieves convergence at $N_s=1.5\times10^5$ but begins to overfit thereafter. Overfitting occurs when a model learns the training dataset too well, including its random fluctuations and statistical noise, increasing generalization error when applied to new data. In this case, the optimal model is achieved for  $N_s=1.5\times10^5$. Notably, this estimate aligns with the estimate obtained from the learning curve of the DNN-DNS model, suggesting that the underlying complexity of the problem being learned is equivalent. To further mitigate the performance bias in DNN models, it may be beneficial to explore the impact of varying model complexity or utilizing optimization algorithms such as Bayesian optimization. However, this requires further investigation and remains a topic for future studies. For the present study, the current model is deemed suitable as it demonstrates satisfactory performance in comparison to the filtered DNS, as discussed in Section~\ref{subsec: DNN validation}.
\subsection{DNN Model Validation} \label{subsec: DNN validation}
Preliminary DNN model validation is performed by obtaining Favre mean, variance, and fourth-order central moments of the $\phi$ obtained from the DNN-FDF models for a validation dataset. These moments are then contrasted against those of the target FDFs extracted from the DNS. Furthermore, to facilitate comparative assessment, the moments from the $\beta$-FDF are also computed and included in the analysis. \textcolor{black}{The results are presented in Figs.~(\ref{DNN_validation_Moments_prediction}, \ref{DNN_validation_fourth_Moments_prediction}) where each symbol represents a single data sample and linear regression line with its coefficient of determination ($R^2$) is included to illustrate the bias in the predictions relative to the filtered DNS data. The correlation coefficient ($r$) is also provided as a measure of the dispersion of the predicted data, with values closer to $1$ indicating a stronger correlation to the filtered DNS data. As displayed in Fig.~(\ref{DNN_validation_Moments_prediction}), both DNN models accurately predict the first moments, as evidenced by the close alignment between the linear regression lines and the ideal $45^{\circ}$ line, along with $r$ and $R^2$ being close to unity.} The $\beta$-FDF model exhibits similar behavior for intermediate $\phi$ values but displays a small bias for extreme $\phi$ values at both ends. Similar observations are made for the second moment, except that the second moment predicted by the $\beta$-FDF deviates further for higher variance values. Both DNN models exhibit a relatively smaller scatter around the mean indicating an overall better correlation with the DNS data than $\beta$-FDF. The fourth order moment as shown in Fig.~(\ref{DNN_validation_fourth_Moments_prediction}) further evaluates the predictive capabilities of these models beyond their input parameters ({\it i.e.}, the first two moments). All models exhibit an increase in bias and dispersion for higher order moments as anticipated. This observed behavior is attributed to the approximated nature of the FDFs derived from these models whose differences with the actual FDFs become increasingly more evident at higher-order moments. All models show a reasonably good correlation with filtered DNS.
\textcolor{black}{The $\beta$-FDF model, despite having slightly higher correlation and determination coefficients, shows a non-linear trend experiencing stagnation at high values of fourth-order moments with a higher level of error.} This suggests that the tails of $\beta$-FDF at such values are consistently under-predicted compared to filtered DNS, consistent with the similar behavior with the variance as observed in Fig.~(\ref{DNN_validation_Moments_prediction}). In contrast, the DNN models exhibit a more even spread out of the fourth order moment with a relatively smaller overall scatter around the DNS values, suggesting a better agreement with the DNS data for this moment. Furthermore, to evaluate the ability of the models to predict the actual shapes of the FDF, Fig.~(\ref{DNN_validation_Sample_FDF}) compares randomly selected predicted FDFs with target FDFs obtained from the DNS as explained in Section \ref{subsec: PPDF_data}. As evidenced by the lower values of JSD, Fig.~(\ref{DNN_validation_Sample_FDF}a, b) demonstrate that all the models accurately predict FDFs which resemble $\delta$ function and Gaussian distribution representing unmixed and well-mixed reactants, respectively. However, when predicting multi-modal FDFs, the DNN models outperform the $\beta$ distribution markedly, as shown in Fig.~(\ref{DNN_validation_Sample_FDF}c, d). DNN models can effectively predict complex FDF shapes for the same input feature space complexity, in contrast to the $\beta$ model. These observations are consistent with previous studies \cite{Tong1402171, Floyd2009} which have also indicated the limitations of the $\beta$-FDF in representing FDFs with complex shapes. 

Overall, the findings obtained from the training and validation process using constant density mixing layer data highlight the strengths of DNN-FDF models in accurately predicting the first and second moments, outperforming the $\beta$-FDF model in terms of bias and dispersion. Although the fourth-order moment introduces some challenges for all models, the DNN models exhibit a more favorable performance overall, with a better correlation to the DNS data. Furthermore, DNN models also exhibit a strong capacity to predict diverse FDF shapes that occur in turbulent flows. To further evaluate the reliability and robustness of the models, we proceed to investigate their performance and generalizability when applied to different mixing layers, as described in the subsequent section. 
\section{Application to Temporal Mixing Layer} \label{sec: results_mixing_layer}
The flow configuration used in this study to evaluate the performance of the DNN-FDF is a non-reacting, 3-D, temporally developing mixing layer. The temporal mixing layer is formed by two parallel streams moving in opposite directions with equal velocities, in a cubic box with the spatial coordinates $x$, $y$, and $z$ representing the streamwise, cross-stream, and spanwise directions, respectively. The cubic box dimensions are $0 \leq x/L_{r} \leq L$, $-L / 2 \leq y/L_{r} \leq L / 2$ and $0 \leq z/L_{r} \leq L$, where $L=L_{v} / L_{r}$ and $L_{r}$ denotes the reference length, as defined below. The length $L_{v}$ is selected such that $L_{v}=2^{N_v \lambda_{u}}$, where $N_v$ is the number of desired successive vortex pairings, and $\lambda_{u}$ is the wavelength of the most unstable mode corresponding to the mean streamwise velocity profile at the initial time. The filtered streamwise velocity, scalar and temperature fields are initialized with hyperbolic tangent profiles subject to free-stream conditions, where the mean streamwise velocity $\langle u\rangle_{L}$ and scalar $\langle\phi\rangle_{L}$ are $\langle u\rangle_{L}=1$ and $\langle\phi\rangle_{L}=1$ on the top, and $\langle u\rangle_{L}=-1$ and $\langle\phi\rangle_{L}=0$ on the bottom. The study considers several density ratios defined as $s=\rho_{2} / \rho_{1}$, where $\rho_{1}$ and $\rho_{2}$ denote the $\langle\rho\rangle_{\ell}$ on the top and bottom free streams, respectively. With a uniform initial pressure field, the initial $\langle T\rangle_{L}$ field is set equal to the inverse of $\langle\rho\rangle_{\ell}$ field based on the ideal-gas equation of state. The flow variables are normalized with respect to the half initial vorticity thickness $L_{r}=\frac12 \delta_{v}(t=0)$ where $\delta_{v}=\Delta U /\left|\partial \overline{\langle u\rangle_{L}} / \partial y\right|_{\max }$ and $\Delta U$ is the velocity difference across the layer; $\overline{(\,)}$ denotes Reynolds-averaged quantities which are constructed from the instantaneous data by spatial averaging over homogeneous ($x$ and $z$) directions. The reference velocity is $U_{r}=\Delta U / 2$ and the reference time is $t_{r}=L_{r}/U_{r}$. The Reynolds number based on the reference values for this simulation is $\mathrm{Re}=U_{r} L_{r} / \nu=50$. The formation of large-scale structures is facilitated by using initial perturbations based on eigenfunctions, resulting in the formation of two successive vortex pairings and strong three-dimensionality. The periodic boundary condition is used in the streamwise and spanwise directions and the zero-derivative boundary condition is used at cross-stream boundaries. The simulations were conducted on equally spaced grid points, with a grid spacing of $\Delta x=\Delta y=\Delta z=\Delta$ and $193^{3}$ and $33^{3}$ grid points for DNS and LES, respectively. The LES filter size is set $\Delta_{f}=2 \Delta$. In DNS, a tophat function is used to filter the data on the same filter size. For LES, two types of simulations are performed, as explained in Section~\ref{sec: formulation}: LES-MC and LES-FD. All FD specifications of LES-FD and LES-MC are similar, except that LES-MC is based on hybrid FD/MC simulations in which an ensemble of Lagrangian particles are randomly distributed throughout the domain. The particle initialization and boundary treatment were made consistent with the LES-FD simulations. There are $6$, $48$ and $384$ particles per grid point for ensemble domain sizes equal to $2\Delta$ ($\Delta_E=2$), $\Delta$ ($\Delta_E=1$) and $\Delta/2$ ($\Delta_E=0.5$), respectively, within the domain at all times, similar to previous work~\cite{Sheikhi2003}. In the constant density case, particles have unity weights but for variable density simulations, their weights are specified initially proportional to $\langle\rho\rangle_{\ell}$ values at each computational cell according to Eq.~(\ref{EQ:rho_FDF}). Additional details on the numerical specifications can be found in the works of Sheikhi~\emph{et al.}~\cite{Sheikhi2003, Sheikhi2007}.

The main objectives of these studies are threefold. Firstly, a consistency check of the DNN-FDF models is performed by comparing the model predictions to those of the FD and the well-established approach of the $\beta$-FDF, as detailed in Section~\ref{subsec: consistency_dnn_fdf}. Secondly, the accuracy of the DNN-FDF model is assessed against the DNS data, as discussed in Section~\ref{subsec: comparison_DNS}. Thirdly, the DNN-FDF model predictions are compared with MC simulations to contrast the methodical differences in obtaining statistical quantities, as presented in \ref{subsec: comparison_MC}. Following a comprehensive comparison of the DNN-FDF models with other methods, Section~\ref{subsec: variable_density} and  Section~\ref{subsec: filter_sdrl} focus on two important aspects of using DNN-FDF models in reacting flows: application to variable-density flow and filtering non-linear variables, respectively. Finally, Section~\ref{subsec: computational_time} sheds light on the average computational requirements for various simulation techniques used in this study. Overall, these objectives are devised to evaluate the performance and capabilities of the DNN-FDF models and their potential applications in turbulent reacting flows.

\textcolor{black}{The summary of governing equations and abbreviations for all simulation methodologies under consideration in this study are listed in Table~\ref{Simulation_method_abbreviations}.} 
\begin{table}[htbp] 
\caption{List of abbreviations for the simulation methodologies for mixing layer simulations}
\begin{tabular}{@{}llll@{}}
\toprule
Simulation methodology      & Scalar FDF/moments solution approach                                        & Governing equations & Abbreviation  \\ \midrule
Direct numerical simulation & ---                                                        & Eqs.~(\ref{Governing_equation})             & DNS           \\
Large eddy simulation       & ---                                                        & Eqs.~(\ref{LES_filterd_equations})             & LES        \\
Large eddy simulation                            & Solution of the first two scalar moments by FD                                                       & Eqs.~(\ref{LES_filterd_equations}, \ref{filterd_mean_equations}, \ref{variance_equations})             & LES-FD        \\
Large eddy simulation                            & Transported FDF with Monte Carlo                           & Eqs.~(\ref{LES_filterd_equations}, \ref{particle_position_eqn}, \ref{particle_scalar_eqn})             & LES-MC            \\
Large eddy simulation                            & DNN model trained on filtered DNS data                       & Eqs.~(\ref{LES_filterd_equations}, \ref{filterd_mean_equations}, \ref{variance_equations}, \ref{softmax})             & DNN-DNS       \\
Large eddy simulation                            & DNN model trained on 0-D PMSR data                        & Eqs.~(\ref{LES_filterd_equations}, \ref{filterd_mean_equations}, \ref{variance_equations}, \ref{softmax})              & DNN-PMSR      \\
Large eddy simulation                            & Assumed $\beta$ function distribution                                     & Eqs.~(\ref{LES_filterd_equations}, \ref{filterd_mean_equations}, \ref{variance_equations}, \ref{beta_function})                   & $\beta$-FDF         \\ \bottomrule       
\end{tabular}
\label{Simulation_method_abbreviations}
\end{table}
\subsection{Consistency of the DNN-FDF} \label{subsec: consistency_dnn_fdf}
\begin{figure}[htbp] 
\centerline{\includegraphics[width = \columnwidth]{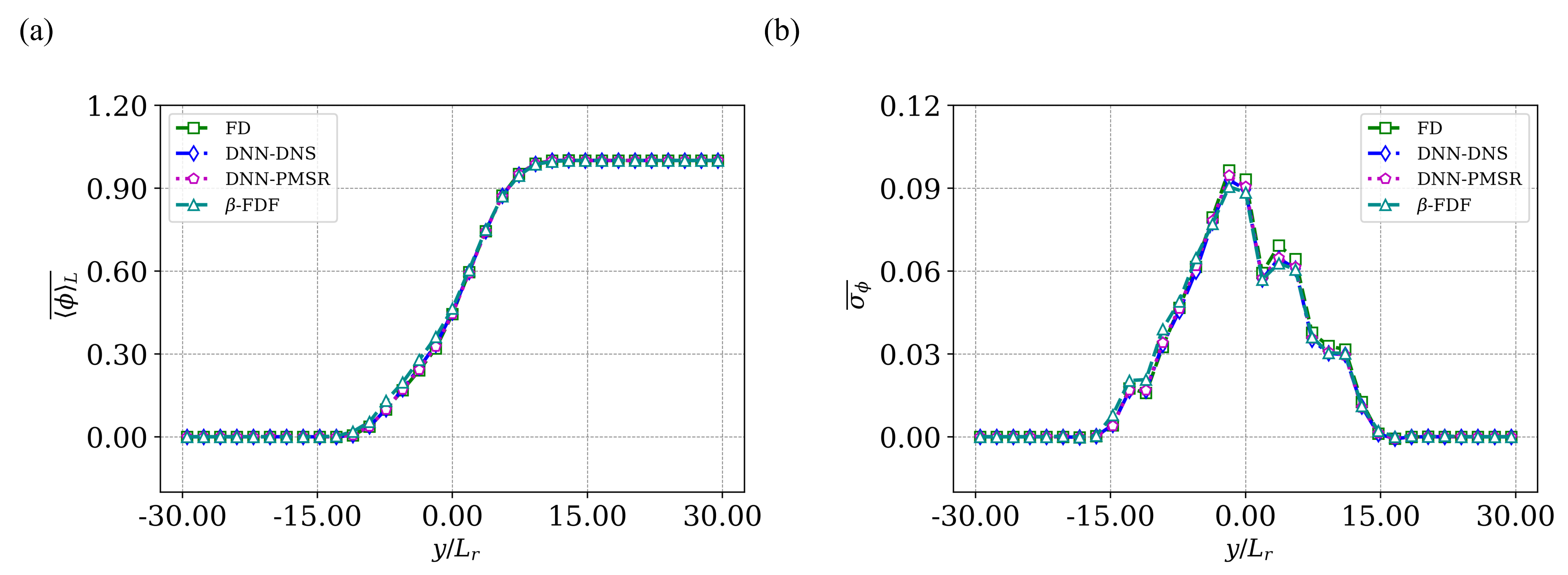}}
\caption{Consistency of the FDF models with respect to FD as demonstrated by comparison of cross-stream variation of Reynolds-average values of (a) ${\langle\phi\rangle_L}$ and (b) ${\sigma_{\phi}}$ in a 3-D temporal mixing layer at $t/t_{r} = 80$ for $s=1$, $C_{\Omega}=1.0$.}
\label{Consistency_FDF_models}
\end{figure}


In this section, the results pertaining to consistency and accuracy assessments of the DNN-FDF methods are presented. The consistency assessment is performed by comparing the first two scalar moments resulting from DNN-FDF with those obtained directly by solving their filtered transport Eqs.~((\ref{filterd_mean_equations}, \ref{variance_equations})) using the FD method. Considering the well-established accuracy of the FD method, this check offers an effective way of evaluating the accuracy of model predictions. Initially, for a broader perspective, Reynolds-averaged statistics are examined. Figure~(\ref{Consistency_FDF_models}) illustrates a comparison between the Reynolds-averaged filtered mean $\phi$ and its variance for constant density flow at $t/t_{r} = 80$. By this time, the flow experiences pairing events and demonstrate significant 3-D effects. To avoid non-realizable FD input values, the presumed FDFs are defined for the region $(1-\langle\phi\rangle_L)\langle\phi\rangle_L/\sigma_\phi > 0$ and treated as delta functions elsewhere. As shown in Fig.~(\ref{Consistency_FDF_models}), the DNN-FDF models accurately predict the mean and variance statistics compared to FD indicating the accuracy of the neural network in predicting these moments. A similar agreement is obtained at other times. The $\beta$-FDF shows similar performance with slight overestimation in both the moments in the bottom region of the mixing layer corresponding to negative $y/L_r$ values. Further assessments are carried out by analyzing instantaneous scatter plots of scalar statistics, as presented in Section~\ref{subsec: variable_density}. Overall, these assessments establish the consistency of the DNN-FDF models in predicting the first two filtered moments.

\subsection{Comparison with DNS} \label{subsec: comparison_DNS}
\begin{figure}[htbp] 
\centerline{\includegraphics[width = \columnwidth]{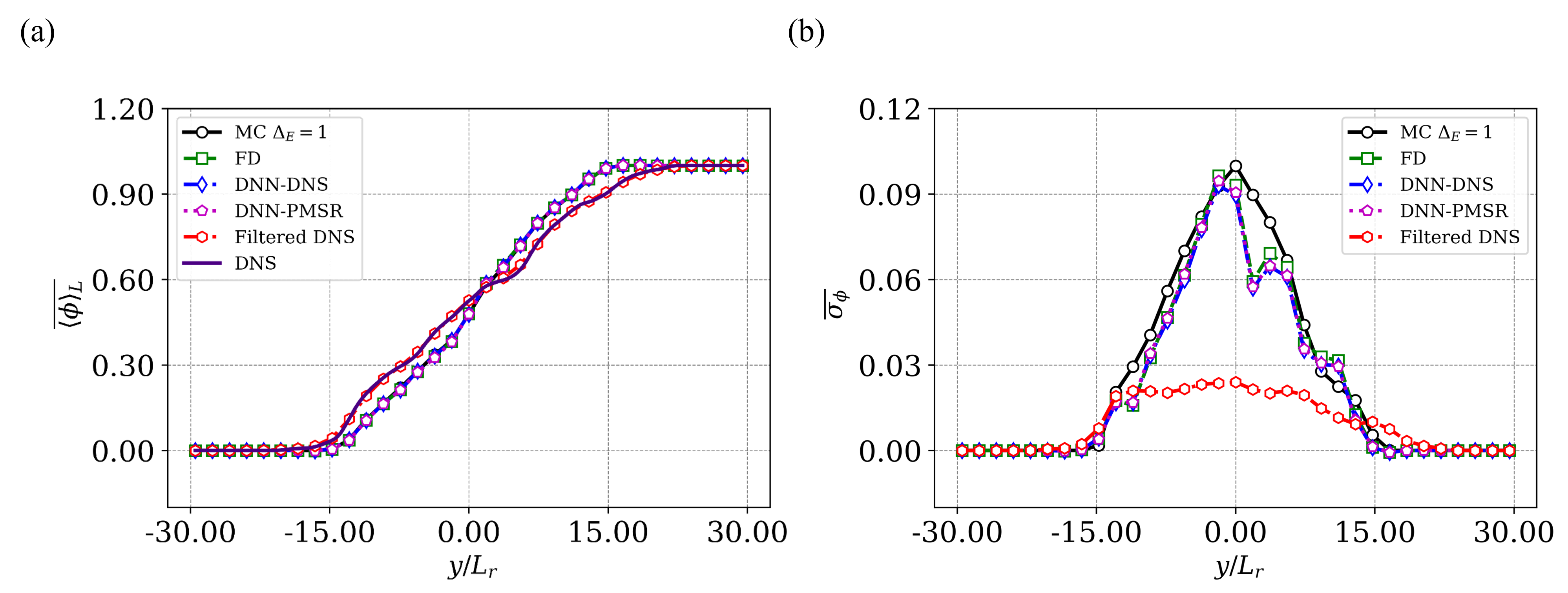}}
\caption{\textcolor{black}{Comparison of cross-stream variation of Reynolds average values of ${\langle\phi\rangle_L}$ and ${\sigma_{\phi}}$ in a 3-D temporal mixing layer as predicted DNN-FDF models and DNS at $t/t_{r} = 80$ for $s=1$, $C_{\Omega}=1.0$.}}
\label{Comparison_DNS_FDF_models_Cphi_1}
\end{figure}

\begin{figure}[htbp] 
\centerline{\includegraphics[width = 0.5\columnwidth]{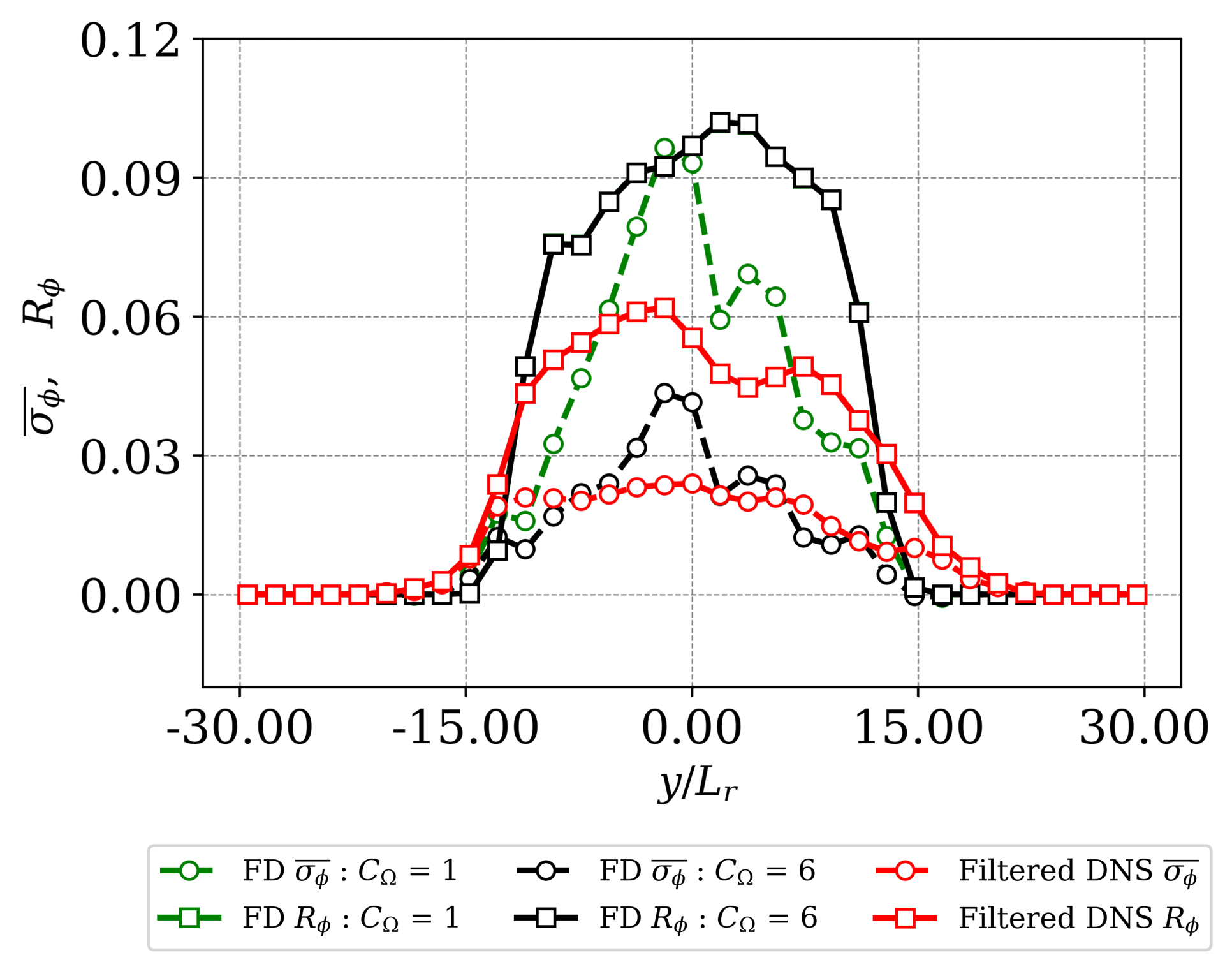}}
\caption{ Resolved ($R_{\phi}$) and SGS (${\sigma_{\phi}}$) components of total scalar variance in a 3-D temporal mixing layer with $s=1$ at $t/t_{r} = 80$ for $C_{\Omega}=1.0$, and $C_{\Omega}=6.0$.}
\label{Composition_of_var_cphi}
\end{figure}

\begin{figure}[htbp] 
\centerline{\includegraphics[width = \columnwidth]{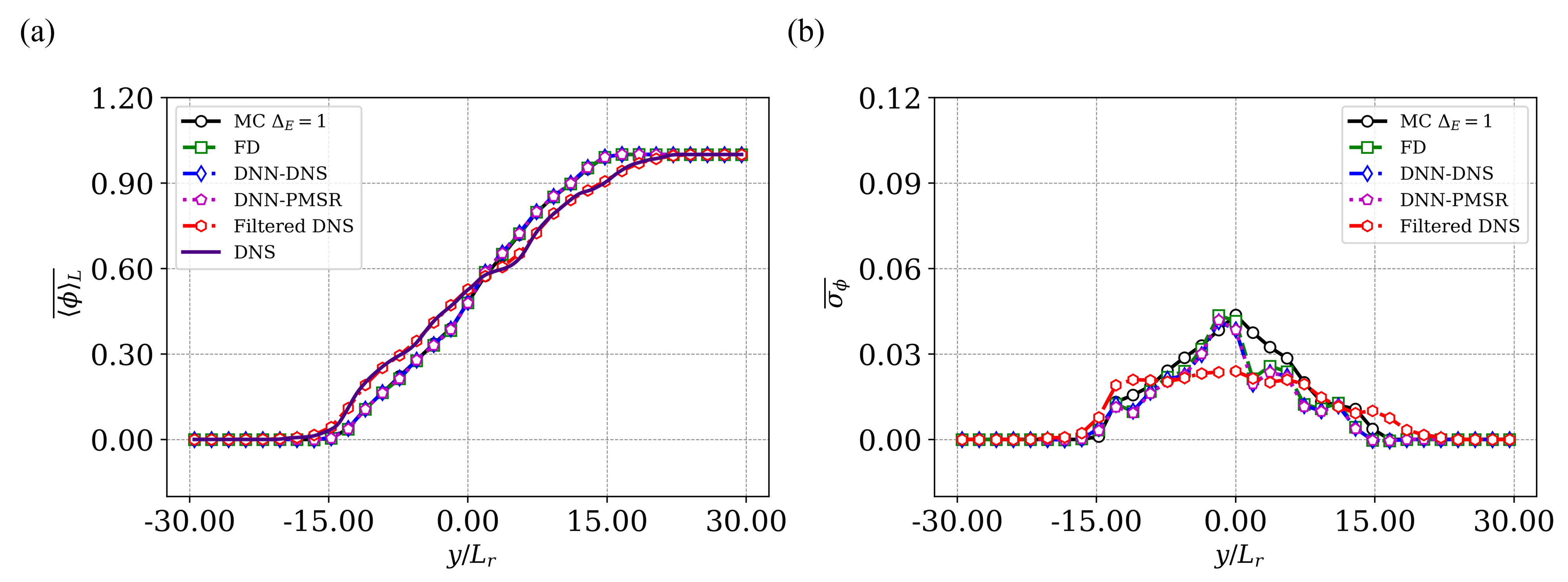}}
\caption{\textcolor{black}{Comparison of cross-stream variation of Reynolds average values of ${\langle\phi\rangle_L}$ and ${\sigma_{\phi}}$ in a 3-D temporal mixing layer as predicted DNN-FDF models and DNS at $t/t_{r} = 80$ for $s=1$, $C_{\Omega}=6.0$.}}
\label{Comparison_DNS_FDF_models_Cphi_6}
\end{figure}
The predictive performance of DNN-FDF models is evaluated through comparative analysis against DNS data. Figure~(\ref{Comparison_DNS_FDF_models_Cphi_1}), presents a comparison between the Reynolds-averaged filtered mean $\phi$ and its variance for constant density flow and the mixing model constant $C_\Omega=1$ at $t/t_{r} = 80$. The mean statistics predicted by DNN-FDF models are in good agreement with the filtered DNS data, as shown in Fig.~(\ref{Comparison_DNS_FDF_models_Cphi_1}a). The accuracy of the filtering operation is substantiated by a perfect match between the filtered and unfiltered DNS results in Fig.~(\ref{Comparison_DNS_FDF_models_Cphi_1}a), which is expected for the tophat filter function. It is evidenced in Fig.~(\ref{Comparison_DNS_FDF_models_Cphi_1}b) that the scalar variance predicted by the DNN-FDF compared well with that from FD but all LES predictions are overestimated relative to DNS. This is not a limitation of the DNN-FDF and it is important to emphasize that DNN-FDF does not bear $C_\Omega$ as a model parameter--- the output of the DNN-FDF is adjusted according to mean and variance input parameters; thus, it is the over-prediction of variance by FD that causes the discrepancy of DNN-FDF with DNS as seen in Fig.~(\ref{Comparison_DNS_FDF_models_Cphi_1}b). The level of variance values here is an artifact of the SGS mixing model and is controlled by the model parameter $C_\Omega$. With $C_\Omega=1$, the FD experiences weak SGS mixing relative to DNS (hence, insufficient scalar dissipation) due to a small mixing model constant value. The scalar variance in LES predictions is obtained from LES-FD (Eq.~(\ref{variance_equations})) wherein the strength of the SGS mixing is manifested in the scalar dissipation term which is proportional to $C_\Omega$. Similarly, in LES-MC formulation, $C_\Omega$ appears in the mixing model (Eq.~(\ref{particle_scalar_eqn})) as a model constant that controls the extent of SGS mixing. To investigate this issue, we examine the SGS and resolved components of the scalar variance for different $C_\Omega$ values ($C_\Omega=1$ and $C_\Omega=6$) obtained from LES-FD as shown in Fig.(\ref{Composition_of_var_cphi}). The resolved variance $R_\phi=\overline{\langle\phi\rangle_{L}\langle\phi\rangle_{L}}-\overline{\langle\phi\rangle_{L}}\ \overline{\langle\phi\rangle_{L}}$ is the part of the total variance $\overline{\phi'^2}$ (where $\phi'=\phi-\overline{\phi}$) resolved by LES  which is unaffected by the SGS mixing (Eq.~(\ref{filterd_mean_equations})) as evident from Fig.(\ref{Composition_of_var_cphi}). Increasing the $C_\Omega$ value leads to higher SGS mixing intensity and larger scalar dissipation rate causing reduced SGS variance by higher dissipation rate of scalar energy at the SGS and hence, decreased total scalar variance. As depicted in Fig.(\ref{Composition_of_var_cphi}), for $C_\Omega=1$, the SGS variance is not only overestimated compared to DNS results but also elevated to the same order as the resolved part which is undesirable in LES. For $C_\Omega=6$, however, the SGS variance is reduced lower than the resolved component, resulting in a better agreement with the DNS which indicates the proper rate of scalar dissipation. The difference between the resolved variance predicted by LES and DNS is related to the LES model closure of the scalar flux and it is irrespective of the SGS mixing model employed. It is worth noting that dependence of LES results with mixing model constant is widely recognized and taken into consideration in several studies; previous investigations have suggested $C_\Omega$ values ranging from $1$ to $8$\cite{Jaberi1999, Colucci1998, Sheikhi2005, Sheikhi2003}, consistent with our findings in the present study. 

Following this analysis, we use $C_\Omega=6$ for all our subsequent LES simulations. \textcolor{black}{Consequently, Fig.~(\ref{Comparison_DNS_FDF_models_Cphi_6}) presents a comparison of statistics for $C_\Omega=6$, where the mean profiles exhibit similar trends to those for $C_\Omega=1$, and all LES simulations display an improved variance prediction compared to filtered DNS.} As shown, both the peak and spread of the scalar variance are predicted well by LES models. These tests highlight the robustness of DNN-FDF and its ability to produce accurate output that solely depends on the accuracy of the input variables ({\it i.e.}, the first two filtered scalar moments obtained from FD).
\subsection{Comparison with LES-MC} \label{subsec: comparison_MC}

\begin{figure}[htbp] 
\centerline{\includegraphics[width = \columnwidth]{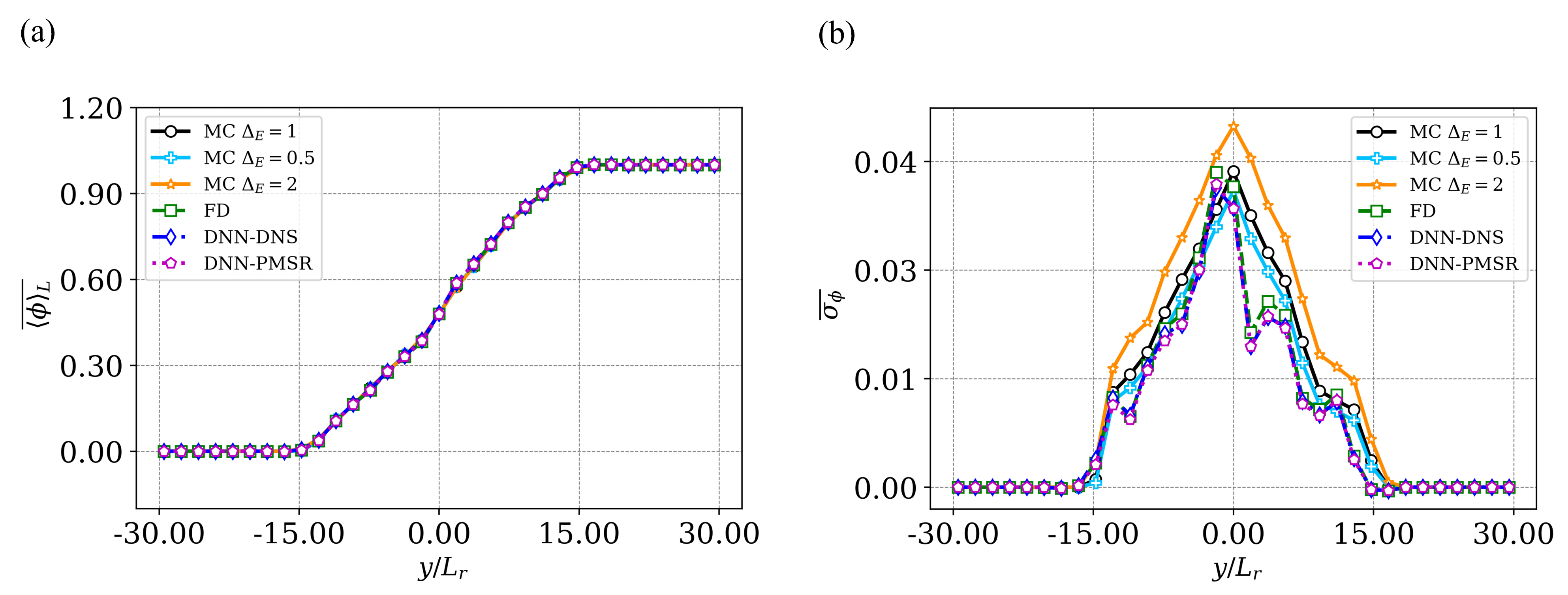}}
\caption{Comparison of DNN-FDF models with MC of cross-stream variation of Reynolds average values of (a) ${\langle\phi\rangle_L}$ and (b) ${\sigma_{\phi}}$ in a 3-D temporal mixing layer for ensemble domain sizes $\Delta_E = 0.5, 1, 2$ at $t/t_{r} = 80$ for $s=1$, $C_{\Omega}=6.0$.}
\label{Comparison_MC_FDF_models}
\end{figure}

\begin{figure}[htbp] 
\centerline{\includegraphics[width = \columnwidth]{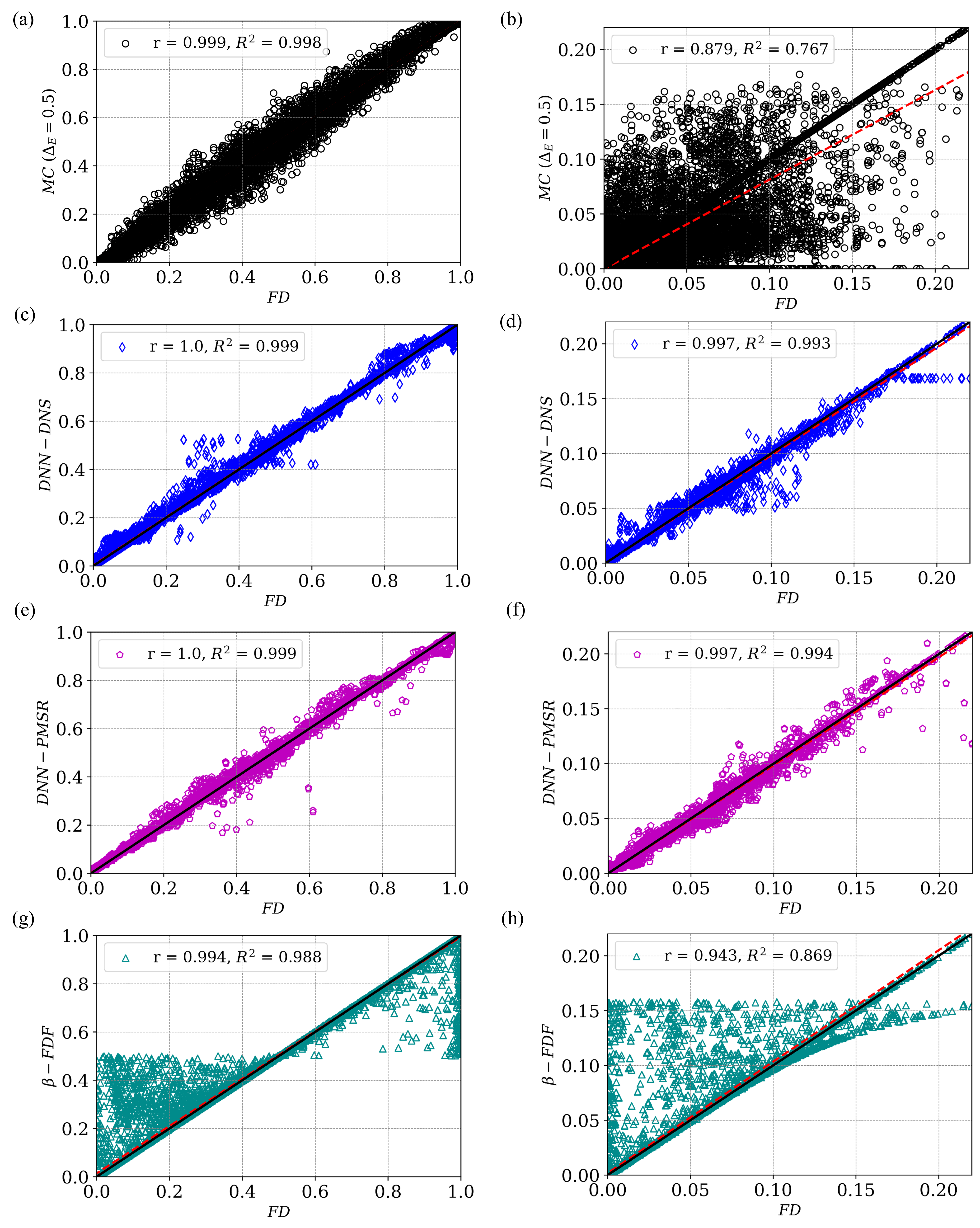}}
\caption{\textcolor{black}{Scatter plots of scalar statistics ${\langle\phi\rangle_L}$ (column 1) and ${\sigma_{\phi}}$ (column 2) in 3-D temporal mixing layer simulations with $s=1$, $C_{\Omega}=6.0$ at  $t/t_{r} = 80$. Comparison of FD with (a,b) MC ($\Delta_E = 0.5$), (c,d) DNN-DNS model, (e,f) DNN-PMSR model, and (g,h) $\beta$-FDF model. The solid and dashed lines denote the linear regression and $45^\circ$ lines, respectively. $r$ denotes the correlation coefficient, $R^2$ denotes the coefficient of determination.}}
\label{Scatter_MC_moments_Cphi_6_Dr_1}
\end{figure}
To further evaluate the performance of the DNN-FDF models, in this section, we compare their predictions with those of the LES-MC as the accuracy of this methodology is well established in previous studies \cite{Colucci1998, Jaberi1999, Sheikhi2003, Sheikhi2005, Sheikhi2007}. 
In LES-MC, the ensemble-averaged quantities are constructed at each FD grid point inside an ensemble domain according to Eq.~(\ref{Q_ensemble_mean}) as explained in Section~\ref{sec: formulation}. To display the extent of variation of LES-MC results with ensemble domain size, we perform simulations with three different sizes $\Delta_E = 0.5, 1, 2$ ({\it i.e.}, ensemble domain sizes of $\Delta/2$, $\Delta$ and $2\Delta$, respectively), while keeping the number of particles within the ensemble domain constant for statistical convergence. The scalar statistics obtained from MC for these cases are compared with DNN-FDF models in Fig~(\ref{Comparison_MC_FDF_models}) for constant density flow and mixing constant $C_\Omega=6$ at $t/t_{r} = 80$. The comparison reveals that the first filtered moment of all ensemble domain sizes agrees well with those obtained by FD and DNN-FDF models, even for large $\Delta_E$ values. The scalar variance (Fig.~(\ref{Comparison_MC_FDF_models}b)), however, shows dependency to $\Delta_E$. The peak value corresponding to MC results decreases with smaller $\Delta_E$ values and it converges to that of the FD, as expected. The closest agreement between MC and FD results is with $\Delta_E=0.5$, although $\Delta_E=1$ also provides reasonably good agreement with lower computational cost. Decreasing $\Delta_E$ in MC increases the computational cost due to the larger number of total particles required to obtain reliable statistics within the smaller ensemble domain. It is evidenced in Fig.~(\ref{Comparison_MC_FDF_models}b) that DNN-FDF provides the closest agreement with the FD results. This highlights an advantage of DNN-FDF in providing results whose accuracy matches that of the FD inputs without requiring any additional parameter such as the ensemble domain size. 

A more thorough comparison of DNN-FDF and MC predictions is by analyzing the instantaneous scatter plots of scalar statistics for constant density flow and mixing constant $C_\Omega=6$ at $t/t_{r} = 80$ in Fig.(\ref{Scatter_MC_moments_Cphi_6_Dr_1}). \textcolor{black}{The results show that both DNN-FDF models and MC (with $\Delta_E=0.5$) accurately predict the first moment, as indicated by the close agreement of their linear regression lines with the $45^{\circ}$ line, and $r$ and $R^2$ being close to unity. For the second moment, both DNN-FDF models exhibit a good correlation with the FD, with low bias and dispersion. However, the MC method shows increased statistical variations for the second moment, resulting in decreased $r$, in accord with previous studies~\cite{Sheikhi2007}.} These variations are symmetric and tend to cancel out, leading to better correlation in terms of Reynolds-averaged quantities, as demonstrated in Fig.~(\ref{Comparison_MC_FDF_models}b). It should be emphasized that an exhaustive investigation of the MC method is outside the scope of the current investigation, and MC results are solely presented for comparative purposes with the DNN-FDF models. Figure~(\ref{Scatter_MC_moments_Cphi_6_Dr_1}) also 
exhibits the superior performance of DNN-FDF models compared to $\beta$-FDF. \textcolor{black}{The regression line for the $\beta$-FDF means shows considerable scatter at extreme $\phi$ values extending to intermediate $\phi$ values. The variance obtained from this model also shows a large scatter overall with increasing bias as variance increases resulting in lower $R^2$.} This is evidenced by the presence of outliers consistently overestimating the local variance in Fig.~(\ref{Scatter_MC_moments_Cphi_6_Dr_1}h). This leads to a slight overestimation of Reynolds-averaged variance as shown in Fig.~(\ref{Consistency_FDF_models}).


\subsection{Variable Density Mixing Layer} \label{subsec: variable_density}
\begin{figure}[htbp] 
\centerline{\includegraphics[width = \columnwidth]{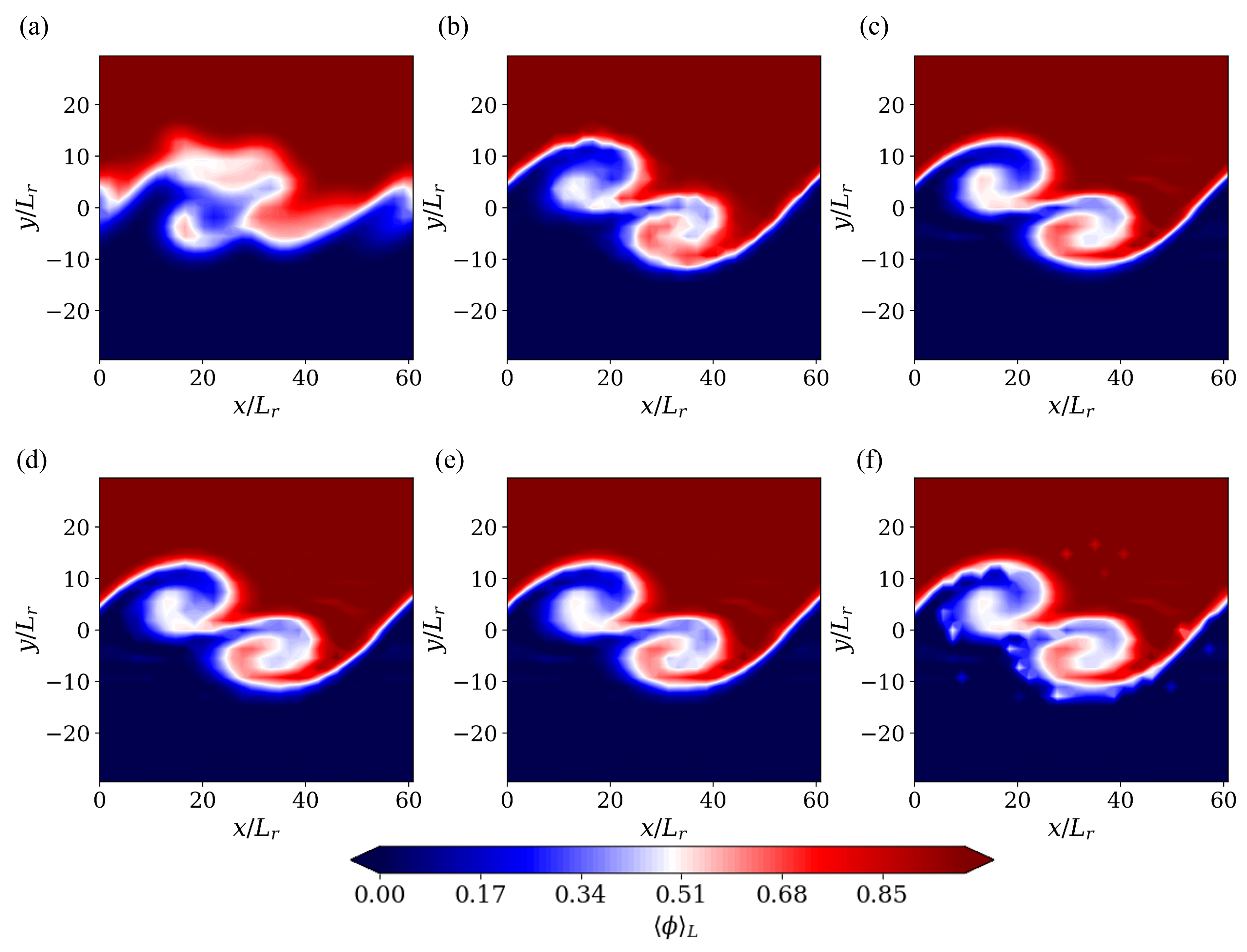}}
\caption{Instantaneous profiles of the $\langle \phi \rangle_L$ on a spanwise plane at $z=0.75L$, $t/t_{r} = 80$ in 3-D temporal mixing layer simulations as obtained for $s=2$, $C_{\Omega}=6.0$; (a) Filtered DNS, (b) MC ($\Delta_E = 1.0$), (c) FD, (d) DNN-DNS model, (e) DNN-PMSR model, and (f) $\beta$-FDF model.}
\label{Scalar_mean_dr_2}
\end{figure}

\begin{figure}[htbp] 
\centerline{\includegraphics[width = \columnwidth]{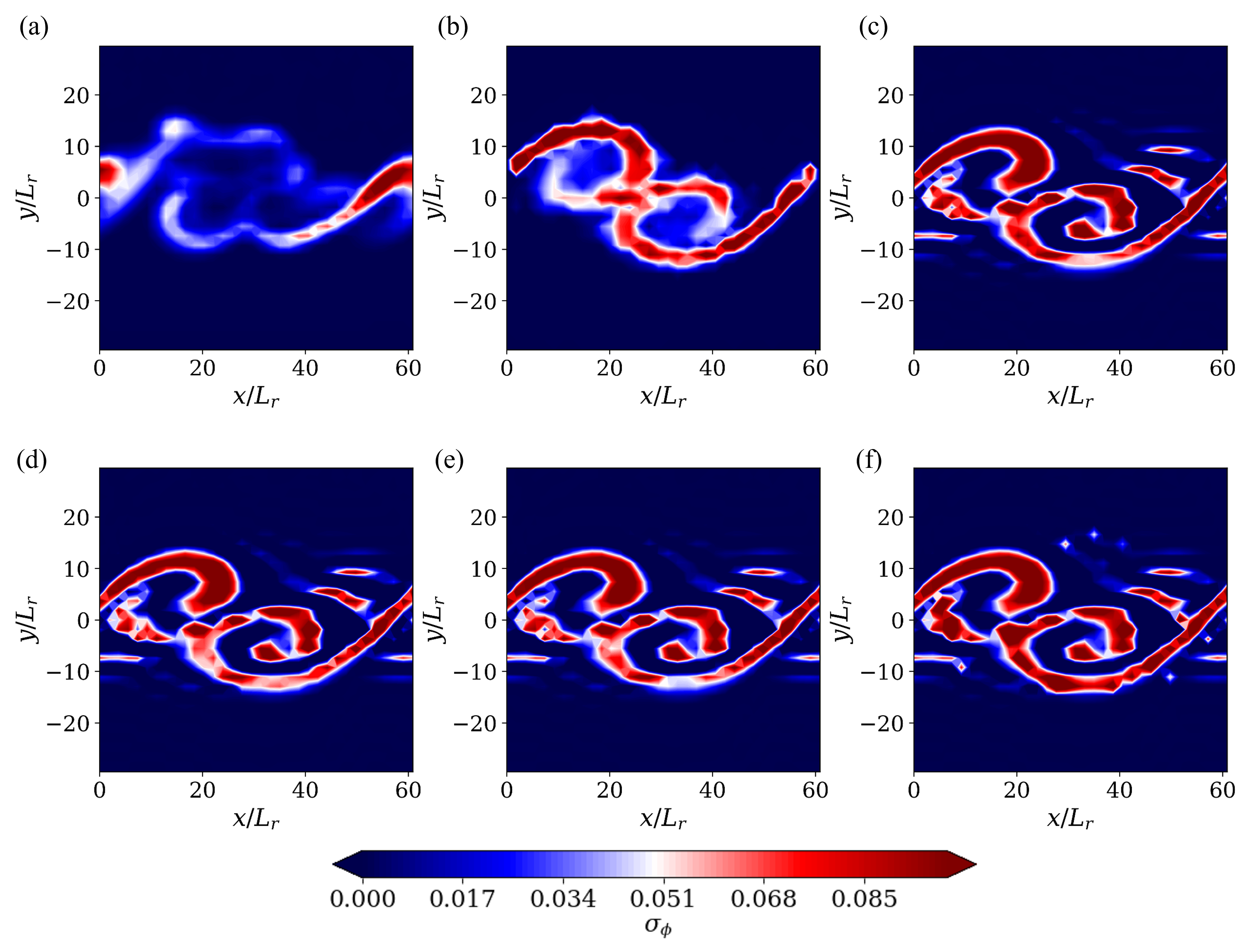}}
\caption{Instantaneous profiles of the $\sigma_{\phi}$ on a spanwise plane at $z=0.75L$, $t/t_{r} = 80$ in 3-D temporal mixing layer simulations as obtained for $s=2$, $C_{\Omega}=6.0$.; (a) Filtered DNS, (b) MC ($\Delta_E = 1.0$), (c) FD, (d) DNN-DNS model, (e) DNN-PMSR model, and (f) $\beta$-FDF model.}
\label{Scalar_var_dr_2}
\end{figure}

\begin{figure}[htbp] 
\centerline{\includegraphics[width = \columnwidth]{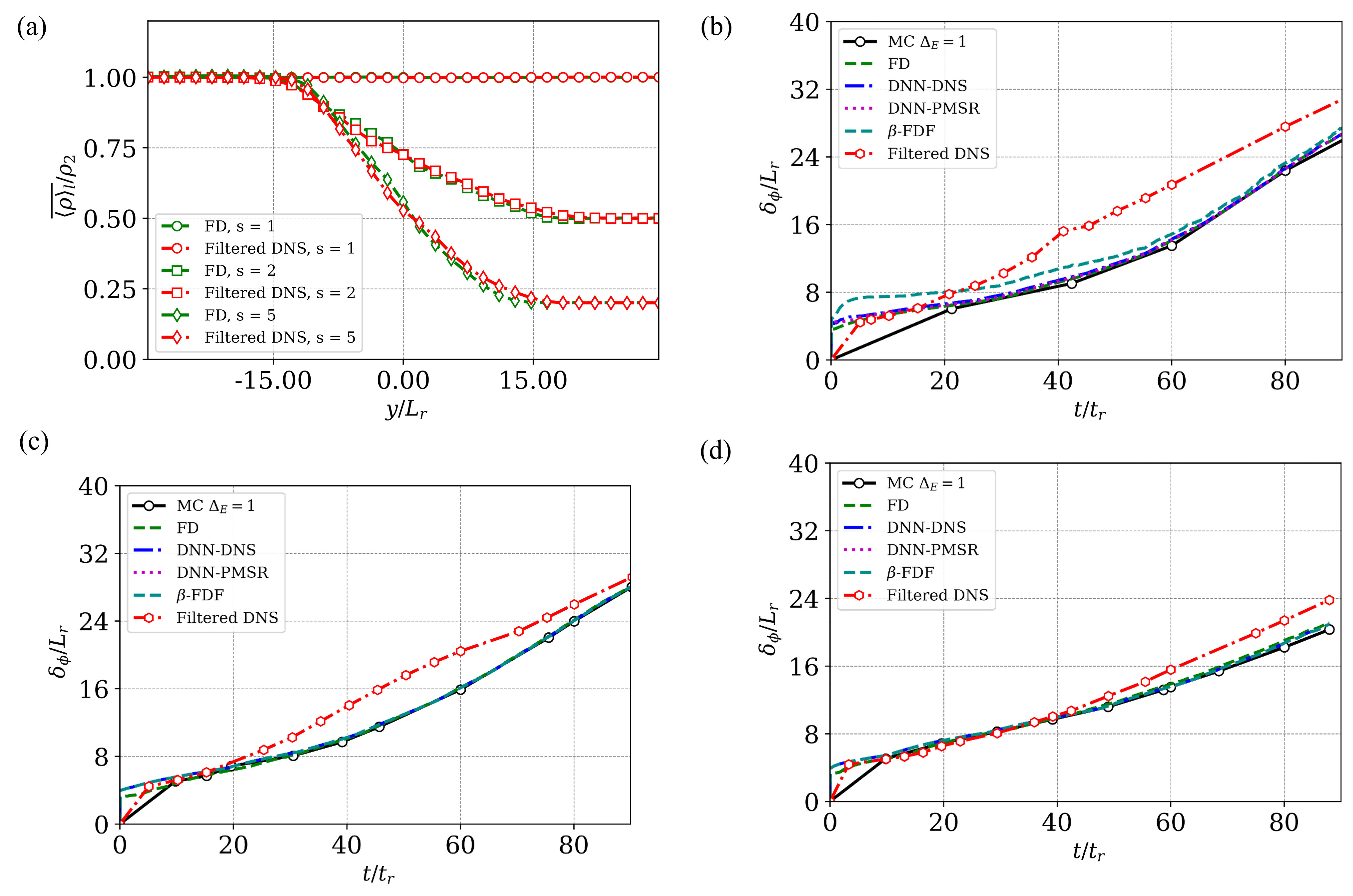}}
\caption{(a) Reynolds averaged profiles of density at $t/t_{r} = 80$ and temporal evolution of scalar thickness in 3-D temporal mixing layer simulations for filtered DNS, MC ($\Delta_E = 1.0$), FD, DNN-DNS model, DNN-PMSR model, and $\beta$-FDF model obtained with (b) $s=1$, (c) $s=2$, and (d) $s=5$.}
\label{scalar_thickness_density}
\end{figure}
The objective of this section is to analyze the performance of DNN-FDF for variable density flows. Simulations are performed of mixing layers with three density ratios, $s=1, 2, 5$ across the layer. For non-unity density ratio cases, the original simulation domain is extended in the $y$ direction without changing the grid resolution to ensure that the zero gradient boundary condition is satisfied when further inhomogeneities are introduced by density variations. Figure~(\ref{scalar_thickness_density}a) shows the Reynolds averaged filtered density variation across the layer for these cases. As shown, for all density ratios LES predictions using FD compare well with the filtered DNS data. This indicates that the FD fields used as input parameters for DNN-FDF are consistent with DNS results. To show the comparative assessments of DNN-FDF for variable density cases, we first examine the instantaneous contours of $\langle \phi \rangle_L$ and $\sigma_{\phi}$ on a spanwise plane at $z=0.75L$ and $t/t_{r} = 80$ as illustrated in Figs.~(\ref{Scalar_mean_dr_2}) and~(\ref{Scalar_var_dr_2}), respectively. These figures show the pairing of two adjacent spanwise rollers at $t/t_{r} = 80$ leading to strong 3-D effects with secondary flow structures on the streamwise planes (not shown). For comparison, the results are presented for all cases: filtered DNS, FD, MC, DNN-FDF models, and $\beta$-FDF. The large scale structures viewed in $\langle \phi \rangle_L$ and  $\sigma_{\phi}$ fields for all LES cases resemble those of filtered DNS. The DNN-FDF model predictions bear a remarkable resemblance to FD, which reaffirms the consistency and accuracy of DNN-FDF similar to the constant density mixing layer. The $\beta$ model reveals more oscillations besides intermittent discontinuity patches in the intense mixing zone and free streams which are in line with the scatter plots in Fig.~\ref{Scatter_MC_moments_Cphi_6_Dr_1}. 
Simulation results using MC generally look similar to those of FD. They however appear to be slightly more oscillating and diffused compared to FD which is due to higher higher level of statistical variations (as shown in Fig.~\ref{Scatter_MC_moments_Cphi_6_Dr_1}) along with the finite ensemble domain size used for ensemble averaging the particles. A similar comparison is observed with $s=5$.
Further, to gain a broader understanding of the model performance for variable density ratios, the scalar thickness ($\delta_{\phi}$) is examined. Scalar thickness is a measure of the thickness of the layer in regard to scalar transport and characterizes the extent of the region where turbulent scalar mixing is in effect. The scalar thickness is defined as
\begin{equation}
\delta_{\phi}(t) = y\left(\overline{\langle\phi\rangle_L} = 0.9\right) - y\left(\overline{\langle\phi\rangle_L} = 0.1\right)
\label{scalar_thickness}
\end{equation}
Figure~(\ref{scalar_thickness_density}b-d) show the time evolution of scalar thickness for all LES models along with filtered DNS for density ratios $s=1$, $2$, and $5$. In general, all LES models exhibit an underprediction of scalar thickness compared to DNS. This is due to the dissipative nature of the Smagorinsky closure which impedes the growth of the layer. This observation is consistent with previous studies~\cite{Sheikhi2003, Sheikhi2007}, which suggest the use of alternate models, such as velocity-scalar filtered density function (VSFDF), to overcome this issue.  The DNN-FDF results are in excellent agreement with FD. The scalar thickness predicted by MC also agrees well with FD with slight statistical variations. The $\beta$ model over-predicts the scalar thickness compared to FD for the constant density case ($s=1$), but its results are in better agreement with other models for larger density ratios. It is worth noting that all models exhibit thinning of the mixing layer with the increasing density ratio, consistent with previous computational~\cite{Sheikhi2007} and experimental~\cite{brown_roshko_1974} studies. These results demonstrate the predictive capabilities of DNN-FDF models trained on constant-density mixing layers for variable-density flows. At low Mach numbers, turbulent mixing is incompressible regardless of the presence of density differences, whether they arise from temperature or molecular weight differences, as shown by Brown \emph{et al.}~\cite{brown_roshko_1974}. Therefore, the DNN models developed on constant-density mixing layers can be applied to low Mach number variable-density turbulent mixing and reacting flows.
\subsection{Filtering of Non-Linear Functions} \label{subsec: filter_sdrl}
\begin{figure}[htbp] 
\centerline{\includegraphics[width = 0.5\columnwidth]{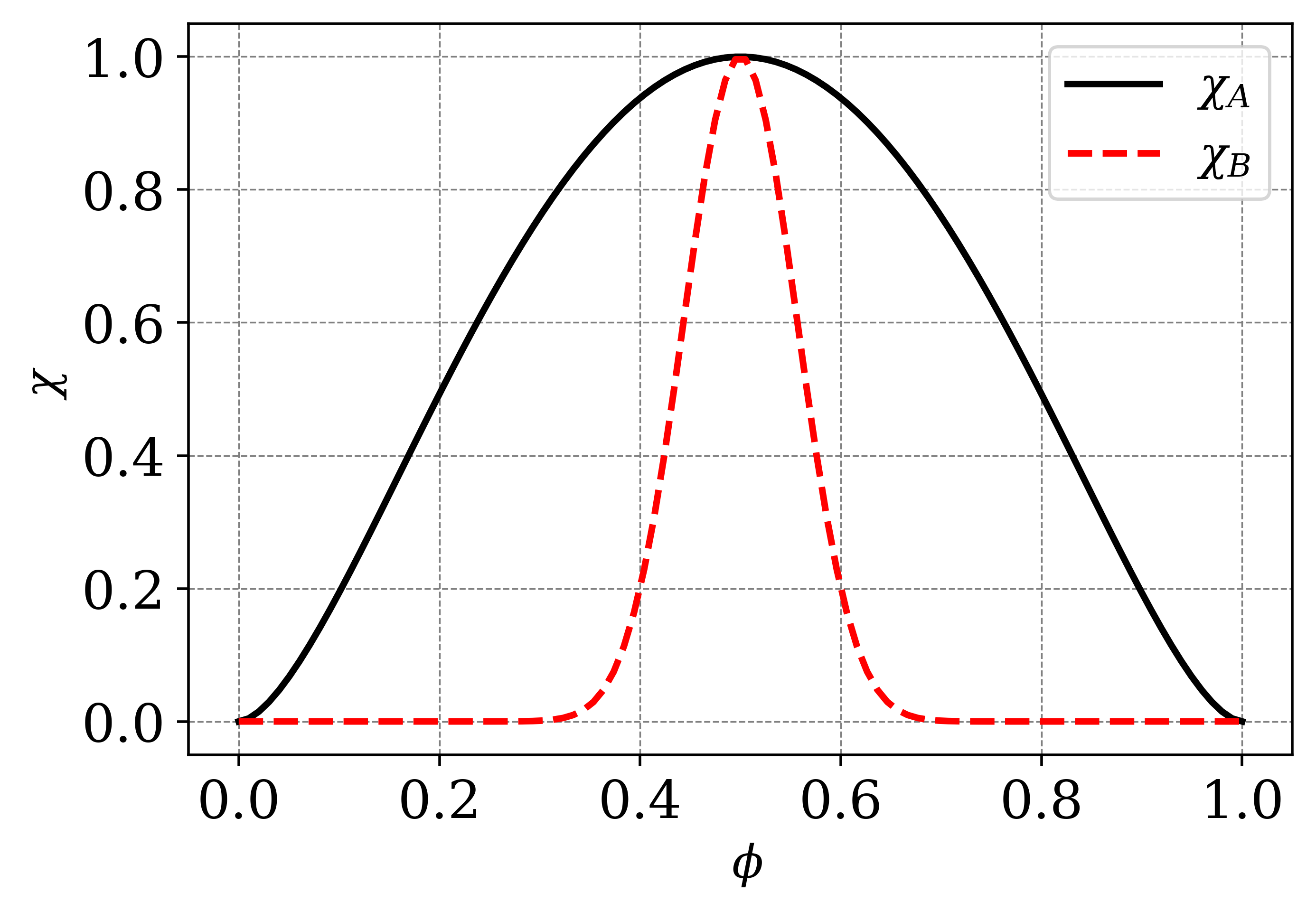}}
\caption{Scalar dissipation profiles $\chi_A$ and $\chi_B$}
\label{SDR_profile}
\end{figure}
\begin{figure}[htbp] 
\centerline{\includegraphics[width = \columnwidth]{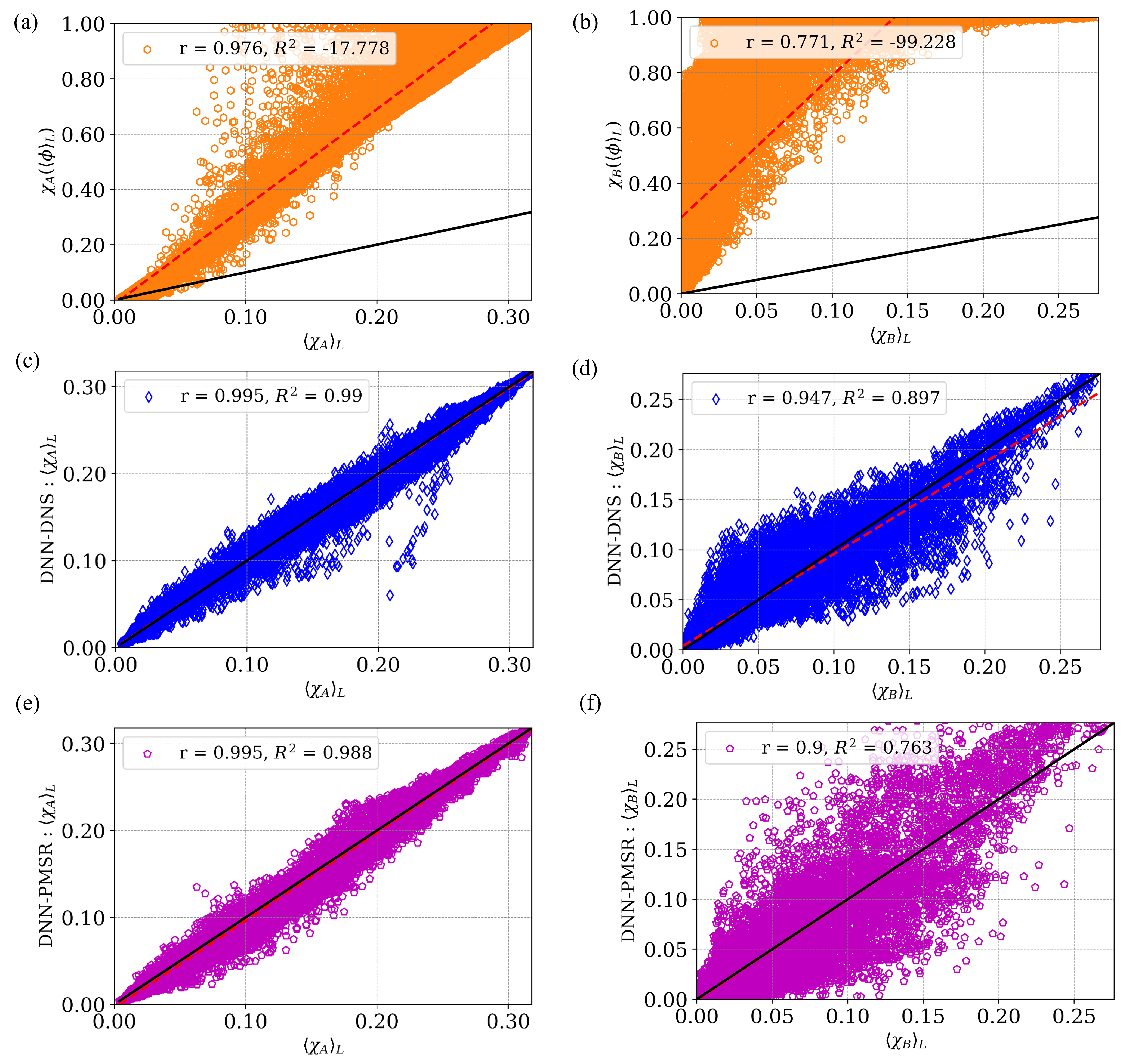}}
\caption{\textcolor{black}{Scatter plots of filtered SDR ${\langle \chi_A \rangle_L}$ (column 1) and ${\langle \chi_B \rangle_L}$ (column 2) in variable density 3-D temporal mixing layer simulation with $s=2$ at collected samples from $t/t_{r}=30,\,60,\,80$. Comparison of filtered DNS with no model (row 1), DNN-DNS model (row 2), and DNN-PMSR (row 3). The solid and dashed lines denote the linear regression and $45^\circ$ lines, respectively. $r$ denotes the correlation coefficient, $R^2$ denotes the coefficient of determination.}}
\label{SDR_scatter_Dr_2}
\end{figure}
A prominent advantage of the FDF methodology is its ability to provide the filtered form of non-linear functions ({\it e.g.}, the chemical reaction source term in reacting flows) without any further modeling assumptions. The capacity of this approach to represent the chemical reaction source term is demonstrated in previous studies~\cite{Jaberi1999, Sheikhi2005, MdG93}. In this section, we evaluate the filtering capability of the DNN-FDF models for non-linear functions. We choose scalar dissipation $\chi$ as a sample non-linear function for this study. The functional expression for $\chi$ in a one-dimensional steady laminar counterflow with $\phi$ ranging from $0$ to $1$ on either side of the reaction zone is given by
\begin{equation} \label{scalar_dissipation_rate}
\chi(\phi) = \text{exp}\{-k[\text{erf}^{-1}(2\phi-1)]^2\}
\end{equation}
where constant $k=2$. The $\chi$ profile generated by this equation is denoted by $\chi_A$ and shown in Fig.~(\ref{SDR_profile}). To assess the filtering performance of FDF models for non-linear functions with sharper gradients, resembling reaction rate functions, we also consider a hypothetical case represented by $\chi_B$ in Fig.~(\ref{SDR_profile}). This hypothetical profile is generated by modifying the constant $k = 50$ in Eq.(\ref{scalar_dissipation_rate}). The resulting filtered values, $\langle\chi\rangle_L$, for these profiles are obtained using Eq.(\ref{TPDF_favre_filtering}) with the DNN-FDF models as well as the ``filtered FDF" generated from DNS data as described in Section \ref{subsec: PPDF_data}. The scatter plots in Fig.(\ref{SDR_scatter_Dr_2}) compare the  DNN-FDF model predictions with those of the filtered FDF for $s=2$. As a reference, Fig~(\ref{SDR_scatter_Dr_2}) also includes the $\langle\chi\rangle_L$ values obtained without any modeling,~\emph{i.e}, using the approximation $\langle\chi(\phi)\rangle_L \approx \chi\left(\langle\phi\rangle_L\right)$, termed ``no model." In the absence of a model, it is observed that the predicted values of $\langle\chi\rangle_L$ for $\langle\chi_A\rangle_L$ are highly overestimated. Additionally, these predictions exhibit a significant dispersion and demonstrate a weak correlation with the filtered DNS predictions. The dispersion and bias become more pronounced for $\langle\chi_B\rangle_L$ due to increased non-linearity. This confirms that the no model approximation is not justified when filtering non-linear functions and proper representation of the SGS variation of scalar becomes necessary. The use of the DNN-FDF model for filtering as presented in Fig~(\ref{SDR_scatter_Dr_2}) shows that this model is able to provide a reasonably accurate prediction of $\langle\chi\rangle_L$.\textcolor{black}{ The accuracy in prediction of $\langle\chi_A\rangle_L$ is evident from the close alignment of the linear regression lines with the $45^{\circ}$ line and correlation/determination coefficient values approaching unity when compared to the filtered DNS data.} Although the correlation slightly weakens for $\langle\chi_A\rangle_B$, resulting in relatively higher bias and variance in the scatter plots, the performance is still considerably better than that of no model, demonstrating the effectiveness of DNN-FDF in approximating filtered quantities dealing with a highly non-linear variation of scalar within the SGS. It is observed that there is higher statistical variation in DNN-PMSR results for $\langle\chi_A\rangle_B$ compared to that of DNN-DNS. This suggests that
further enhancements to the DNN-PMSR model could be achieved by incorporating more representative training samples and properly conditioning the training data. While these observations demonstrate the efficacy of DNN-FDF models in filtering highly non-linear functions, the potential implications of these findings for turbulent reacting flow simulations in practice remain promising and necessitate further investigations in the future.
\subsection{Applicability of DNN-FDF for Varied Grid and Filter Sizes} \label{subsec: appln_grid_filter_size}
\begin{figure}[htbp] 
\centerline{\includegraphics[width = \columnwidth]{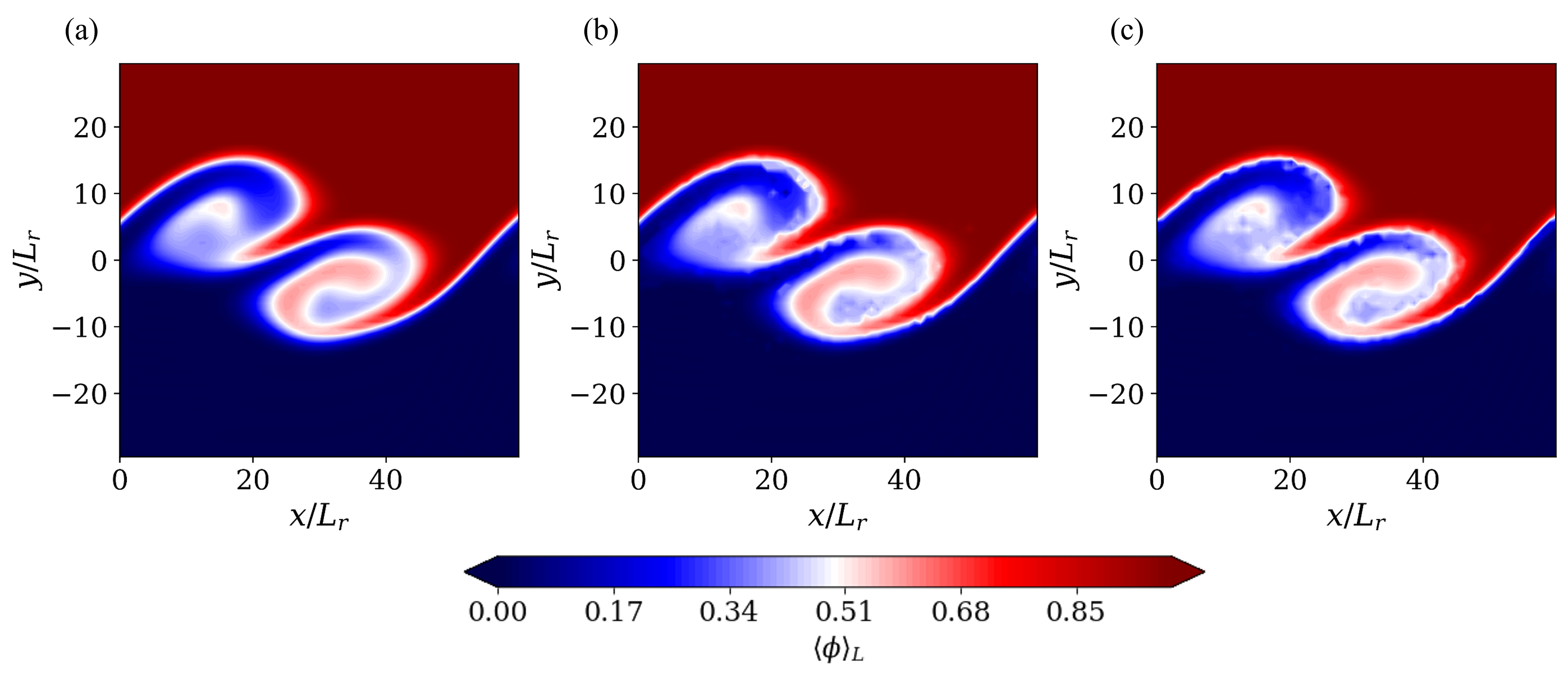}}
\caption{\textcolor{black}{Instantaneous profiles of the $\langle \phi \rangle_L$ on a spanwise plane at $z=0.75L$, $t/t_{r} = 80$ in LES of 3-D temporal mixing layer simulations with $66^{3}$ grid points as obtained for $s=2$ and constant filter size $\Delta_f^I = \Delta_f = 4\Delta^+$; (a) FD, (b) DNN-DNS model, and (c) DNN-PMSR model.}}
\label{high_res_scalar_mean_dr_2_deltaf_4}
\end{figure}
\begin{figure}[htbp] 
\centerline{\includegraphics[width = \columnwidth]{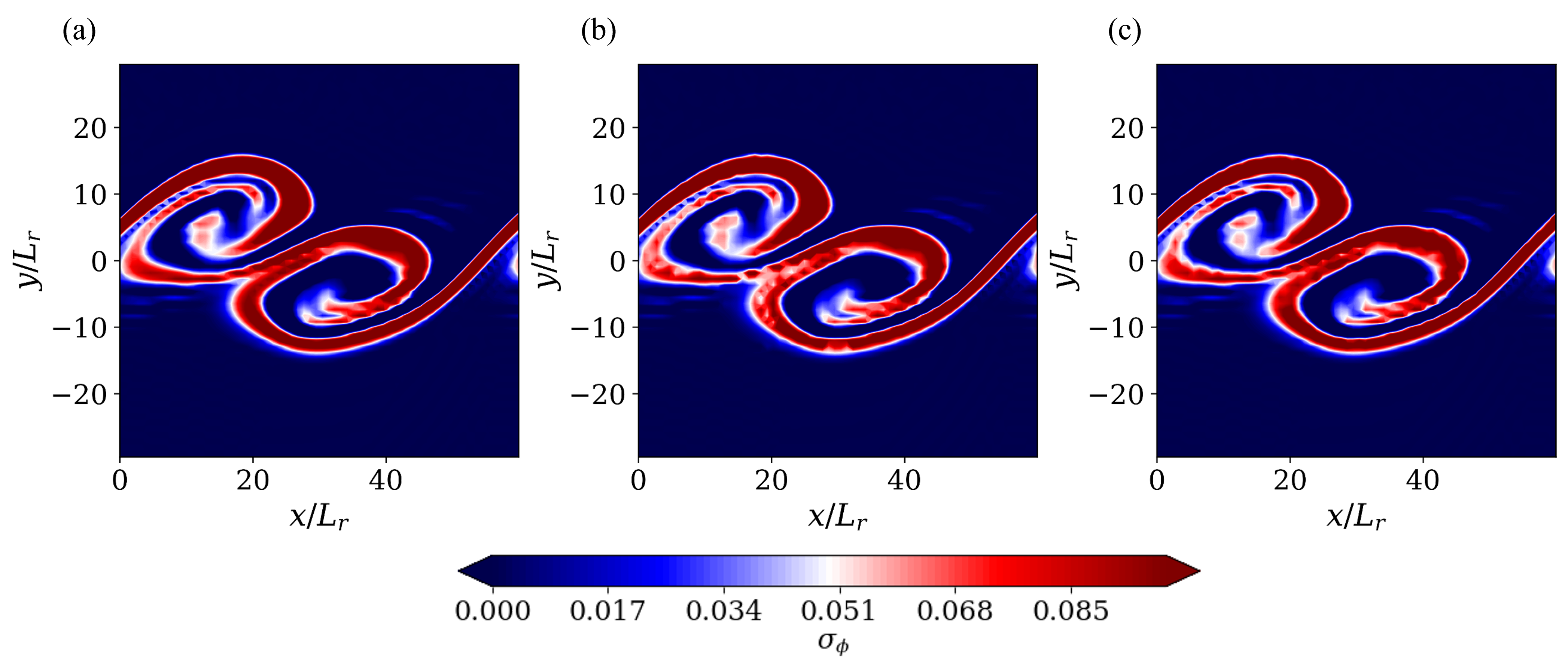}}
\caption{\textcolor{black}{Instantaneous profiles of the $\sigma_{\phi}$ on a spanwise plane at $z=0.75L$, $t/t_{r} = 80$ in  LES of 3-D temporal mixing layer simulations with $66^{3}$ grid points as obtained for $s=2$ and constant filter size $\Delta_f^I = \Delta_f = 4\Delta^+$; (a) FD, (b) DNN-DNS model, and (c) DNN-PMSR model.}}
\label{high_res_scalar_var_dr_2_deltaf_4}
\end{figure}
\begin{figure}[htbp] 
\centerline{\includegraphics[width = \columnwidth]{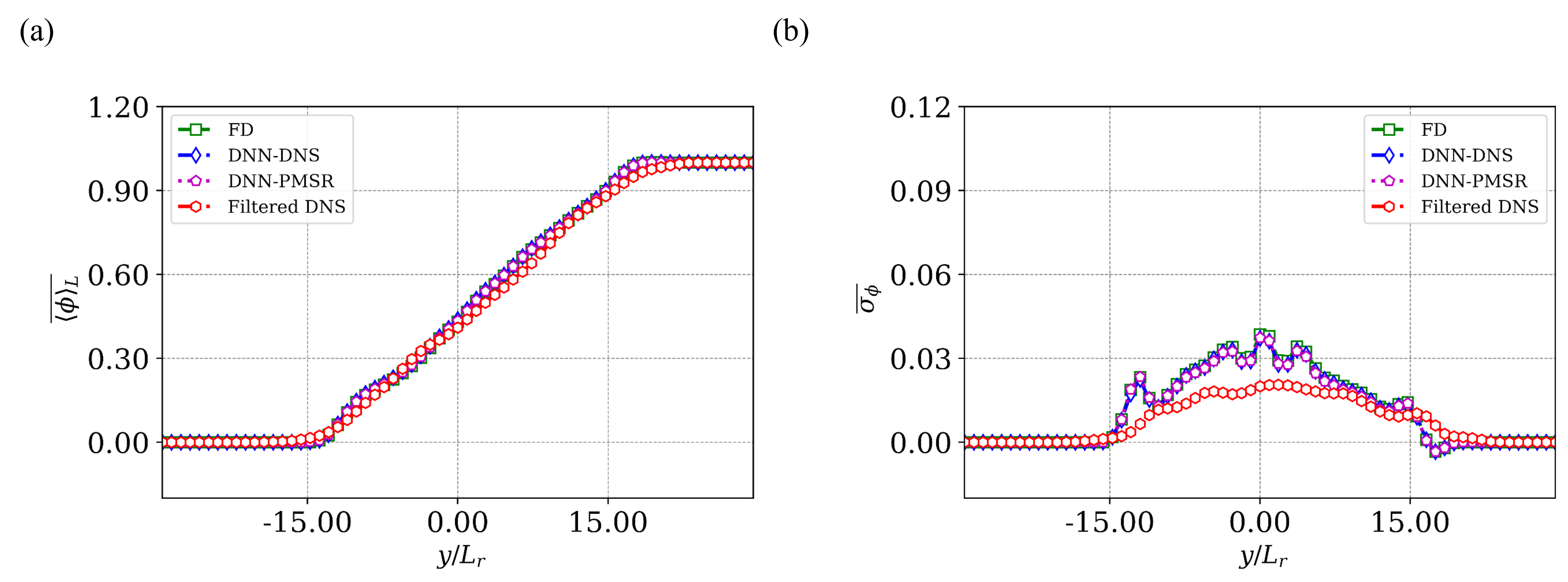}}
\caption{\textcolor{black}{Comparison of cross-stream variation of Reynolds average values of ${\langle\phi\rangle_L}$ and ${\sigma_{\phi}}$ in LES of 3-D temporal mixing layer simulations with $66^{3}$ grid points as predicted DNN-FDF models and DNS at $t/t_{r} = 80$ for $s=2$ and constant filter size $\Delta_f^I = \Delta_f = 4\Delta^+$.}}
\label{Comparison_high_res_moments_deltaf_4}
\end{figure}
\begin{figure}[htbp] 
\centerline{\includegraphics[width = \columnwidth]{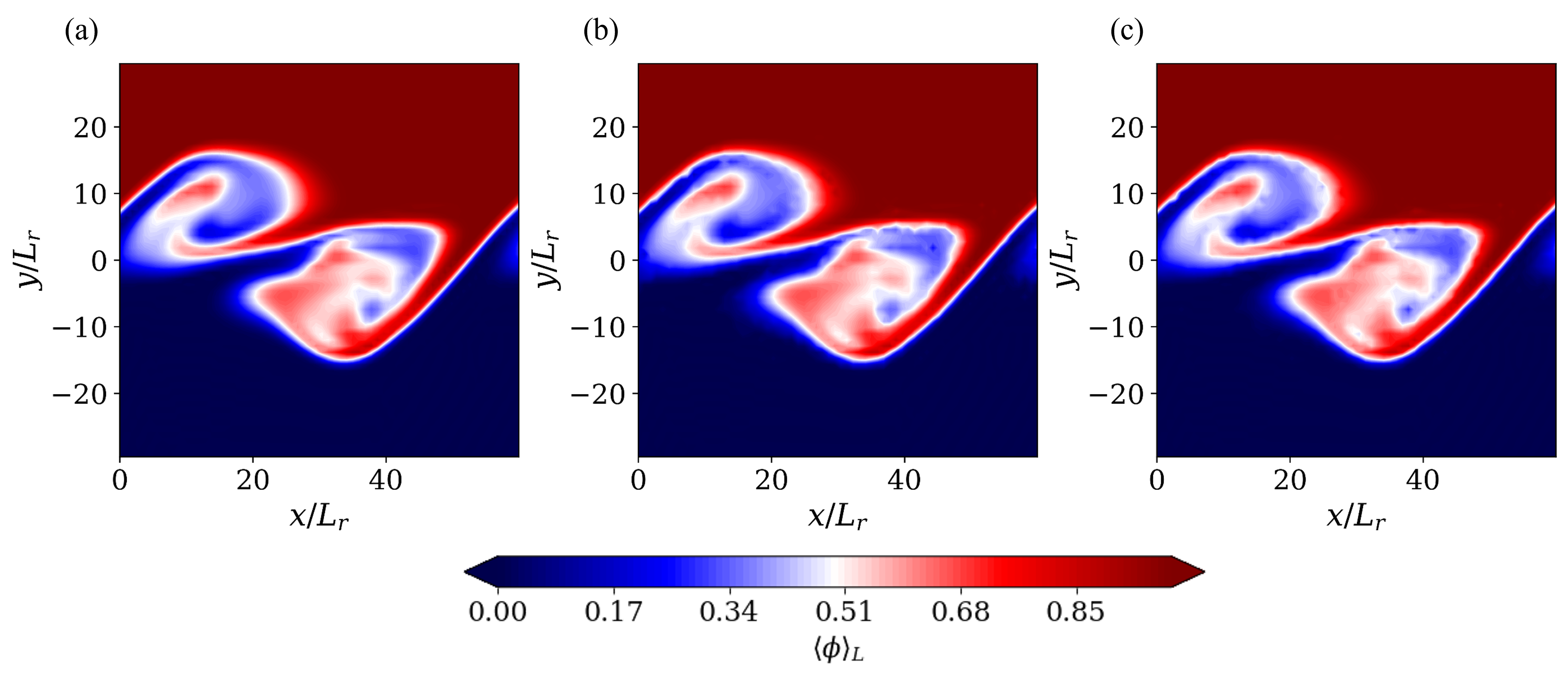}}
\caption{\textcolor{black}{Instantaneous profiles of the $\langle \phi \rangle_L$ on a spanwise plane at $z=0.75L$, $t/t_{r} = 80$ in LES of 3-D temporal mixing layer simulations with $66^{3}$ grid points as obtained for $s=2$ and half filter size $\Delta_f^{II} = \Delta_f/2 = 2\Delta^+$; (a) FD, (b) DNN-DNS model, and (c) DNN-PMSR model.}}
\label{high_res_scalar_mean_dr_2_deltaf_2}
\end{figure}
\begin{figure}[htbp] 
\centerline{\includegraphics[width = \columnwidth]{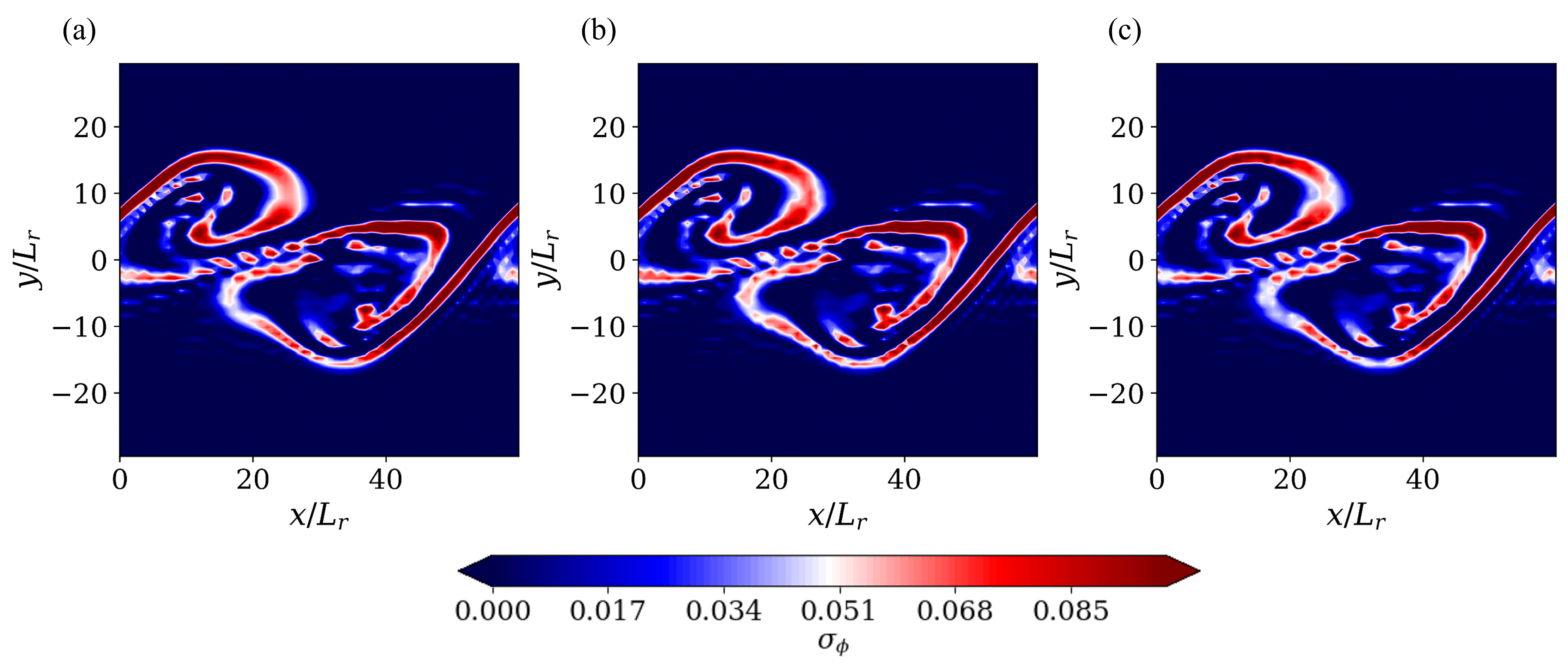}}
\caption{\textcolor{black}{Instantaneous profiles of the $\sigma_{\phi}$ on a spanwise plane at $z=0.75L$, $t/t_{r} = 80$ in  LES of 3-D temporal mixing layer simulations with $66^{3}$ grid points as obtained for $s=2$ and half filter size $\Delta_f^{II} = \Delta_f/2 = 2\Delta^+$; (a) FD, (b) DNN-DNS model, and (c) DNN-PMSR model.}}
\label{high_res_scalar_var_dr_2_deltaf_2}
\end{figure}
\begin{figure}[htbp] 
\centerline{\includegraphics[width = \columnwidth]{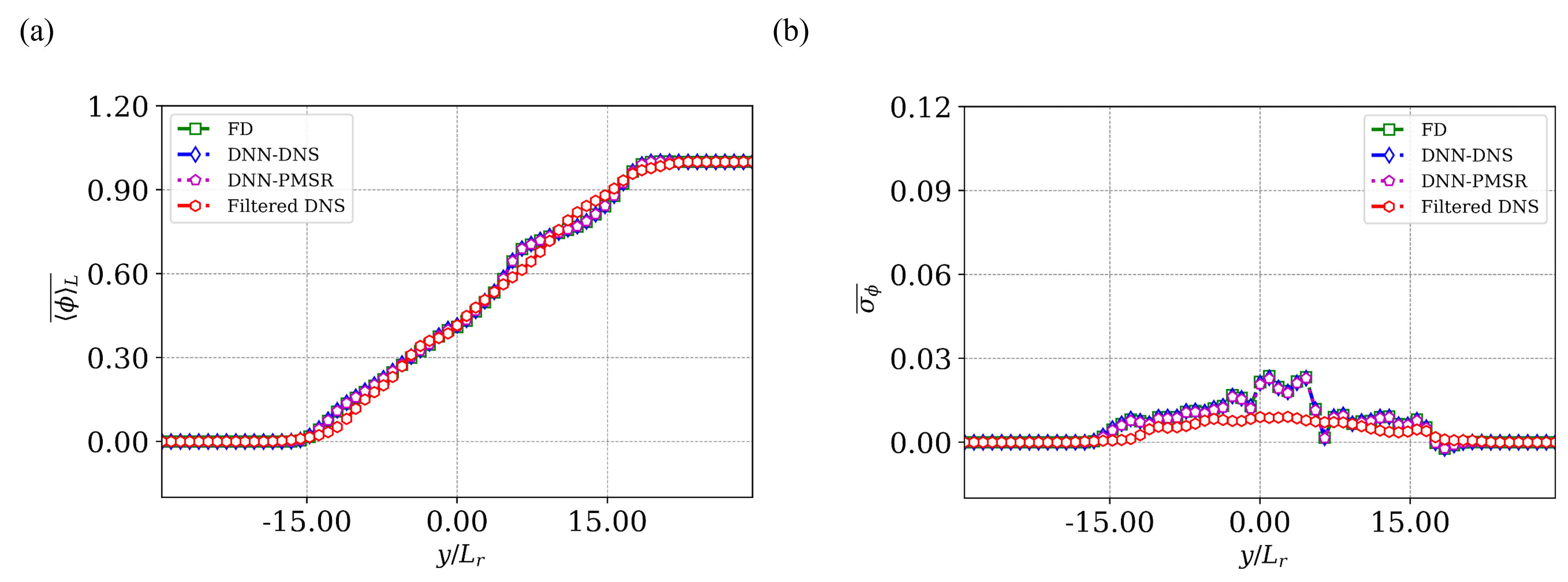}}
\caption{\textcolor{black}{Comparison of cross-stream variation of Reynolds average values of ${\langle\phi\rangle_L}$ and ${\sigma_{\phi}}$ in LES of 3-D temporal mixing layer simulations with $66^{3}$ grid points as predicted DNN-FDF models and DNS at $t/t_{r} = 80$ for $s=2$ and half filter size $\Delta_f^{II} = \Delta_f/2 = 2\Delta^+$.}}
\label{Comparison_high_res_moments_deltaf_2}
\end{figure}
\textcolor{black}{The filter and grid sizes are not explicit parameters in the DNN approach, as the effect of the SGS is embedded within the variance itself. This characteristic is evident in the performance of DNN-PMSR, which is trained on data generated from particle simulations in the pairwise mixing stirred reactor (PMSR) and yet yields satisfactory results when applied to LES of variable density mixing layers. To further demonstrate the versatility and predictive performance of DNN-FDF models using different LES grid resolution and filter sizes, we study two mixing layer cases with $s=2$, each simulated at a higher grid resolution ($\Delta^+=\Delta/2$) with $66^3$ grid points. In case I, the filter size was kept similar to that of the original resolution ($\Delta_f^I = \Delta_f$); this case is focused on the grid resolution effect at a constant filter size. In case II, the filter size is halved ($\Delta_f^{II} = \Delta_f/2$) to reveal the influence of filter size variation with similar grid resolution (of $\Delta^+$).
Figures~(\ref{high_res_scalar_mean_dr_2_deltaf_4}, \ref{high_res_scalar_var_dr_2_deltaf_4}) present instantaneous contours of the mean scalar, $\langle \phi \rangle_L$, and its variance, $\sigma_{\phi}$, on a spanwise plane at $z=0.75L$ and $t/t_{r} = 80$ for case I. These figures illustrate higher resolution prediction of mean and variance by the FD compared to Figs.~(\ref{Scalar_mean_dr_2},\ref{Scalar_var_dr_2}). Remarkably, the DNN-FDF model predictions closely resemble the FD results. Additionally, Fig.~(\ref{Comparison_high_res_moments_deltaf_4}) demonstrates that the Reynolds-averaged statistics obtained with the DNN-FDF model show excellent agreement with the respective filtered DNS and FD results. The comparisons of DNN-FDF model predictions to FD and filtered DNS for case II, exhibit similar trends, as presented in Figs.~(\ref{high_res_scalar_mean_dr_2_deltaf_2}, \ref{high_res_scalar_var_dr_2_deltaf_2}, \ref{Comparison_high_res_moments_deltaf_2}). As expected, the flow structures in Figs.~(\ref{high_res_scalar_mean_dr_2_deltaf_4}, \ref{high_res_scalar_var_dr_2_deltaf_4}) closely resemble those observed in lower LES resolution results from Figs.~(\ref{Scalar_mean_dr_2},\ref{Scalar_var_dr_2}), as the filter size remains the same in both cases. However, for case II with the smaller filter size, finer structures are observed, resulting in a smaller SGS contribution and, consequently, a decrease in variance as indicated by Fig.~(\ref{Comparison_high_res_moments_deltaf_2}). Similar to all cases presented above, it is observed that DNN-FDF consistently provides the same quality of results as the FD.}

\textcolor{black}{ The extensive tests presented in this section demonstrate the robustness and adaptability of DNN-FDF models in accurately representing FDFs with different LES grid resolutions and filter sizes without requiring any modifications. The DNN-FDF models effectively capture the statistical features of the FDF in different cases considered here, affirming their predictive capabilities and generalizability.}
\subsection{Computational Time} \label{subsec: computational_time}
\begin{table}[htbp] 
\caption{Total computational times for the reacting jet simulations}
\begin{tabular}{@{}lllll@{}}
\toprule
Simulation & FDF Model & Grid resolution & Normalized CPU time &  \\ \midrule
LES-FD     & $\beta$-FDF      & $33^3$              & $1$                                             &  \\
           & DNN-DNS   & $33^3$              & $2.9$                                 &  \\
           & DNN-PMSR  & $33^3$              & $2.9$                                 &  \\
LES-MC ($\Delta_E =1$)     &           & $33^3$             & $3.9$                                  &  \\
LES-MC ($\Delta_E =0.5$)     &           & $33^3$             & $29$                                  &  \\
DNS        &           & $193^3$             & $166$                                   &  \\ \bottomrule
\end{tabular}
\label{Normalized_CPU_time}
\end{table}
In order to assess the computational demands of the simulations considered here, the average computational time required is recorded and presented in Table~(\ref{Normalized_CPU_time}). To facilitate the comparison, the simulation times are normalized by the CPU time required for the LES-FD approach with the $\beta$-FDF model. It is observed that the computational time required for the LES-FD simulations using the DNN-FDF model, including the DNN evaluation and moment calculations, is marginally greater than that for the $\beta$-FDF. This indicates that DNN-FDF evaluation and moment calculation is slightly more computationally expensive than evaluating the $\beta$-FDF calculations for each computational cell. The DNN models used in the simulations are trained in \texttt{PyTorch} using \texttt{Python} implementation. However, when utilized for the simulations, trained models are evaluated through PyTorch \texttt{C++ API}, which is observed to be slightly slower than the \texttt{Python} library. This factor could have also contributed to the slight slowdown observed in the speed of DNN-FDF models compared to $\beta$-FDF. When comparing the computational cost of LES-MC simulations, it is evident that these simulations are computationally more demanding than the $\beta$-FDF and DNN-FDF simulations due to the additional demands of MC simulations. This difference becomes even more significant as the ensemble domain size is reduced--- smaller ensemble domain size necessitates a larger number of MC particles to achieve statistical convergence within the ensemble domain. 

Overall, the computational time necessary for LES-FD simulations utilizing the DNN-FDF is marginally larger than that for the $\beta$-FDF model. DNN-FDF however demonstrates higher fidelity in predicting scalar statistics within the SGS, as evidenced in this study through comparison with DNS and LES-MC (even with the smallest ensemble domain size considered). The DNN-FDF can thus provide a cost-effective alternative to $\beta$-FDF to represent the scalar FDF with higher accuracy. This is particularly important for simulations requiring the SGS variation of mixture fraction, {\it e.g.}, reacting flow simulations via a flamelet model~\cite{Sheikhi2005}. As expected, the computational requirement of DNS significantly exceeds that of any LES methodology. LES-MC is more costly than DNN-FDF but it requires a fraction of DNS CPU time. This suggests that LES-MC can be employed to develop cost-effective DNN models for cases where DNS is not feasible. 

\section{Conclusions} \label{sec: conclusions}
In this study, we investigate the performance of the deep neural network (DNN) models for modeling filtered density function (FDF) of mixture fraction in a variable density, three-dimensional (3-D), temporal mixing layer with conserved scalar mixing. First, a systematic method for selecting the training sample size and architecture of the DNN-FDF models is proposed by minimizing bias and variance through the learning curves, which can be used to guide the development of DNN models for other applications. Two DNN models are introduced by training the DNN using the FDF constructed from the DNS data, as well as that generated using a zero-dimensional pairwise mixing stirred reactor (PMSR) case. 
The latter approach is introduced as an alternative means of generating the training data when DNS is not feasible. 
Subsequently, we conduct comprehensive investigations of the consistency and convergence of DNN-FDF models and assess their accuracy against DNS, transported FDF using the Monte Carlo (MC) method, and the widely used presumed $\beta$-FDF approach. The key insights of this comparative analysis are
\begin{enumerate}
    \item Both DNN-FDF models are consistent with FD and have better predictive capabilities over the conventional $\beta$-FDF, particularly for multi-modal FDF shapes and at higher variance values. 
    \item \textcolor{black}{Favorable predictions obtained from the DNN-FDF trained on PMSR data highlight the potential of this computationally economical approach to generate the training data in the absence of DNS results.}    
    \item 
    Subsequent to their development, the accuracy of DNN-FDF models is solely reliant on the quality of their input variables and remains independent of the model constants used in generating these inputs.
    \item The DNN-FDF models and MC approach demonstrate comparable performance in terms of the first two filtered moments. However, ensemble averaging operation in MC depends on the ensemble domain size, which not only affects the accuracy of the results but also influences their computational demand. The DNN-FDF models are independent of any additional parameter for averaging.
    
    \item The DNN-FDF models trained on constant-density mixing layers are readily applicable to low Mach number turbulent flows with variable density. The DNN-FDF model predictions of scalar statistic and scalar thickness compare satisfactorily well with filtered DNS for various density ratios.
    

    \item \textcolor{black}{The developed DNN-FDF models are demonstrated to exhibit adaptability to different grid resolutions and filter sizes. This adaptability is attributed to the fact that the filter and grid sizes are not explicit parameters in the DNN approach, and the effect of the SGS is inherently encapsulated within the input variance data.}
    
    \item \textcolor{black}{The DNN-FDF models provide favorable agreement with filtered DNS in predicting the filtered value of non-linear functions. This demonstrates their predictive capability in accounting for filtered non-linear terms as encountered in, {\it e.g.,} LES of reacting flows, which is an important application of DNN-FDF.} 
    
\end{enumerate}

The findings of this study demonstrate the capabilities of DNN-FDF models in LES-FD simulations as an affordable and reliable approach compared to DNS and LES-MC.
The DNN-FDF models present a more accurate and efficient approach to capturing scalar statistics while only marginally increasing the computational cost in comparison to the widely used $\beta$-FDF model. These promising results illustrate the potential of DNN-FDF models for predicting turbulent reactive flow simulations, particularly at low Mach numbers. Overall, this study establishes the groundwork for further research on utilizing DNN-based models in turbulent reactive flow simulations.
\begin{acknowledgments}
This study is supported in part by the Office of the Vice President for Research at the University of Connecticut through the Research Excellence Program. The authors acknowledge the computing resources provided by the high-performance computing facilities at the University of Connecticut. The authors thank Dr. Farhad Imani for many valuable discussions which have contributed greatly to the quality of this study.
\end{acknowledgments}

\section*{Data Availability Statement}
The data that support the findings of this study are available from the corresponding author upon reasonable request.
\section{References}
\bibliography{main}

\end{document}